\documentclass[pre,preprintnumbers,amsmath,amssymb]{revtex4}
\usepackage{color,hyperref}
\usepackage{graphicx}
\usepackage{epsfig}
\usepackage{epstopdf}
\usepackage{bm}

\newtheorem{theorem}{Theorem}[section]

\DeclareMathOperator{\arcsinh}{arcsinh}

\begin{document}

\title{The closeness of localised structures between the Ablowitz-Ladik lattice and  Discrete Nonlinear Schr\"odinger equations II: Generalised AL and DNLS systems  }
\author{Dirk Hennig}
\author{Nikos I. Karachalios}
\affiliation{Department of Mathematics, University of Thessaly, Lamia, GR 35100, Greece}
\author{Jes\'{u}s Cuevas-Maraver}
\affiliation{Grupo de F\'{i}sica No Lineal, Departamento de F\'{i}sica Aplicada I,
Universidad de Sevilla. Escuela Polit\'{e}cnica Superior, C/ Virgen de \'{A}frica, 7, 41011-Sevilla, Spain \\
Instituto de Matem\'{a}ticas de la Universidad de Sevilla (IMUS). Edificio Celestino Mutis. Avda. Reina Mercedes s/n, 41012-Sevilla, Spain}


\begin{abstract}
The Ablowitz-Ladik system, being one of the few integrable nonlinear lattices, admits a wide class of analytical solutions, ranging from exact spatially localised solitons to rational solutions in the form of the spatiotemporally localised discrete Peregrine soliton. Proving a closeness result between the solutions of the Ablowitz-Ladik and a wide class of Discrete Nonlinear Schr\"odinger systems in a sense of a continuous dependence on their initial data, we establish that such small amplitude  waveforms may be supported in the nonintegrable lattices, for significant large times. The nonintegrable systems exhibiting  such  behavior include a generalisation of the Ablowitz-Ladik system with a power-law nonlinearity and the  Discrete Nonlinear Schr\"odinger with  power-law and saturable nonlinearities. The outcome of numerical simulations illustrates in an excellent agreement with the analytical results the persistence of small amplitude Ablowitz-Ladik analytical solutions in all the nonintegrable systems considered in this work, with the most striking example being that of the Peregine soliton.

\end{abstract}

\maketitle

\section{Introduction}
The Ablowitz-Ladik equation (AL) \cite{AL},\cite{AL2},\cite{Herbst},
\begin{equation}
i\dot{\psi}_n=\kappa\nu(\psi_{n+1}-2\psi_n+\psi_{n-1}) +\mu\,|\psi_{n}|^2(\psi_{n+1}+\psi_{n-1}),\,\,\,n\in {\mathbb{Z}},\label{eq:AL0}
\end{equation}
is the famous integrable discretisation of the cubic nonlinear Schr\"{o}dinger partial differential equation
\begin{eqnarray}
	i\partial_t u + \nu u_{xx}+\gamma |u|^{2}u=0,\;\; x\in\mathbb{R},\label{eq:NLS0}
\end{eqnarray}
The integrability of \eqref{eq:AL0} was proved by the discrete version of the Inverse Scattering Transform \cite{AL}.  The AL is one of the few known completely integrable (infinite) lattice systems  admitting exact soliton solutions \cite{Ablowitz},\cite{Faddeev},\cite{Ver1},\cite{Maruno}. The one-soliton solution of AL reads as
\begin{equation}
\begin{split}
\psi^s_n&=\frac{\sinh\beta}{\sqrt{\mu}}\mathrm{sech}\left[\beta(n-ut)\right]\exp(-i(\omega t-\alpha n)),\\
\;\;\omega&=-2\cos \alpha \cosh \beta,\\
u&=2\beta^{-1}\sin \alpha \sinh \beta,\label{eq:one-soliton}
\end{split}
\end{equation}
with $\alpha \in [-\pi,\pi]$ and $\beta \in [0,\infty)$.
In \cite{akhm_AL},\cite{akhm_AL2}, it was shown that the AL \eqref{eq:AL0}, admits (again in similarity  with the integrable NLS \eqref{eq:NLS0}), rational solutions as the discrete  counterpart of the Peregrine soliton (AL-PS)
\begin{equation}
	\psi_{n}^r(t)=q\left[1-\frac{4(1+q^2)(1+4iq^2t)}{1+4n^2q^2+16q^4t^2(1+q^2)}\right]e^{2iqt}.
	\label{prw_exact}
\end{equation}
where the parameter $q$ fixes a  background amplitude. Such exact solutions do not exist for the with regards to applications in physics  more relevant  non-integrable Discrete Nonlinear Schr\"{o}dinger equation (DNLS)
\begin{equation}
i\dot{\phi}_n+{\kappa\nu}\left(\phi_{n+1}-2\phi_n+\phi_{n-1}\right)+\gamma
|\phi_n|^2\phi_n=0,\,\,\,n\in {\mathbb{Z}},\;\;\gamma\in\mathbb{R}.\label{eq:DNLS0}
\end{equation}
However, in \cite{DNJ2021}, we established the closeness of the solutions of the AL system \eqref{eq:AL0} to the DNLS \eqref{eq:DNLS0} in the $l^2$ and $l^{\infty}$-metrics, in a sense of a continuous dependence on their initial data.
The most significant application of this closeness result is that  the DNLS lattice admits small amplitude solutions of the order ${\cal{O}}(\varepsilon)$ that stay ${\cal{O}}(\varepsilon^3)$-close to the analytical solutions of the AL for any  finite time $0<T_{\small{f}}<\infty$.  Furthermore, taking the advantages offered by the discrete ambient space, we extended the result to higher dimensional lattices and other examples of nonlinearities.  The  numerical studies of \cite{DNJ2021} for the persistence of the AL soliton in the cubic DNLS \eqref{eq:DNLS0} and the saturable one (see below), corroborated to high accuracy the analytical predictions that these non-integrable DNLS models sustain  small amplitude solitary waves preserving for significantly large time intervals the functional form and characteristics (as the velocity) of the analytical AL-one soliton solution.

In the present paper we provide further analytical as well numerical extensions of the results of \cite{DNJ2021}: On the analytical side, we establish the closeness of the solutions between the following generalisation of the Ablowitz-Ladik lattice (G-AL),
\begin{equation}
i\dot{U}_n=\kappa\nu(U_{n+1}-2U_n+U_{n-1}) +\delta\,|U_{n}|^{2\sigma}(U_{n+1}+U_{n-1}),\,\,\,n\in {\mathbb{Z}},\label{eq:AL0s}
\end{equation}
and the integrable AL \eqref{eq:AL0}. Our study in this paper for the G-AL \eqref{eq:AL0s} is motivated by the work in \cite{JGAL} where numerous features were explored ranging  from the derivation of conservation laws of the model, the numerical identification of discrete solitons and the study of their stability- as the nonlinearity exponent $\sigma$ is varied- through extended Vakhitov–Kolokolov criteria, to the presentation of numerical evidence for quasi-collapse as indicated by the appearance of large amplitudes for large values of $\sigma$. The results of \cite{JGAL} justified  that the G-AL \eqref{eq:AL0s} similarly to the other important generalised DNLS models with  power nonlinearity \cite{GDNLS1}, \cite{GDNLS2}, \cite{GDNLS3},  \begin{eqnarray}
	\label{pN}
	F(z)z&=&\gamma |z|^{2\sigma}z,
\end{eqnarray}
are significant  for the study of the transition from integrability  when $\sigma=1$, to non-integrability for $\sigma>1$, and the potential structural changes to the dynamics of intrinsic localised modes possessed by these systems \cite{GDNLS2},\cite{Kim}.  In this regard, we think that our closeness approach  is of particular relevance: it simultaneously establishes the existence and persistence of small amplitude localised waveforms for a general class of  non-integrable DNLS systems which approximate the functional form of the analytical solutions of the integrable AL \eqref{eq:AL0}. As it was mentioned above, the analysis covers also the DNLS with saturable nonlinearity,
\begin{eqnarray}
\label{satn}
F(z)z=\frac{\gamma z}{1+|z|^2},
\end{eqnarray}
a model governing the dynamics of photo-refractive media; we refer to the works \cite{Satn1},\cite{Satn2},\cite{Satn3},\cite{Satn4},\cite{Satn5} concerning the propagation and stability of discrete solitons. Thus, extra motivation is given to pursue further numerical studies exploring the closeness of solutions of the integrable AL to the solutions of the above generalised G-AL and DNLS systems.

The presentation of the paper is as follows: in section \ref{SecII} we prove the closeness of the solutions between the integrable AL \eqref{eq:AL0} and its generalisation \eqref{eq:AL0s}.  The proof follows  the lines in \cite{DNJ2021} aided this time by the nontrivial conserved quantity of the G-AL which involves the Gauss hypergeometric function \cite{JGAL}.  Section \ref{SecIII} is devoted to the numerical studies and is divided into three parts. In the first subsection \ref{GALsol} we present the results of the numerical study concerning the persistence of small amplitude single-soliton solutions \eqref{eq:one-soliton} of the AL in the quartic G-AL \eqref{eq:AL0s} corresponding to the case $\sigma=2$. It is shown that the AL-soliton persists when implemented as an initial condition to the G-AL for significantly large times.   Furthermore, the numerical results for the evolution of the norms of the distance function of the corresponding solutions not only fulfill but prove to be 
  even sharper than the respective analytical estimates. A notable feature is that the evolving soliton in the G-AL lattice preserves with a high accuracy the speed of the analytical AL-soliton solution. The same excellent agreement with the analytical predictions is illustrated in the subsection \ref{qDNLS} for the dynamics of the analytical AL-soliton in the DNLS with the power nonlinearity \eqref{pN} in its quintic case $\sigma=2$.  We also test the limits of the  analytical persistence results regarding the smallness of the amplitudes of the AL-soliton initial conditions in the cubic $\sigma=1$ as well as in the quintic DNLS $\sigma=2$.  While the analytical estimates are again satisfied, we find that the persistence of the AL-soliton in these non-integrable lattices lasts for significantly lower times (in comparison to the smaller amplitude cases); this is in accordance with the analysis, as our closeness results rely, in the notion of continuous dependence, on the initial conditions. Furthermore we observe the emergence of novel dynamical features: in the cubic DNLS the AL-soliton  evolves as a high time-period spatially localised breather, while in the quintic DNLS we observe a self-similar decay of the initial condition. Still in both cases, we observe the preservation of the speed of the initial pulse. The third part in subsection \ref{PerDyn} concludes with the presentation  of a particularly interesting application of the closeness of solutions approach: the persistence of small amplitude rational solutions of the form of the AL-PS \eqref{prw_exact}. Remarkably, and yet in excellent agreement with the analytical considerations, the numerical findings corroborate the existence/persistence of spatiotemporal localised waveforms significantly close to the AL-PS, in all the cases of the non-integrable lattices considered herein. Finally, section \ref{secSum} summarises the main findings and gives an outlook to  extensions for 
  future studies.

%
\begin{figure}[tbp!]
	\begin{center}
		\begin{tabular}{cc}
			\includegraphics[width=.50\textwidth]{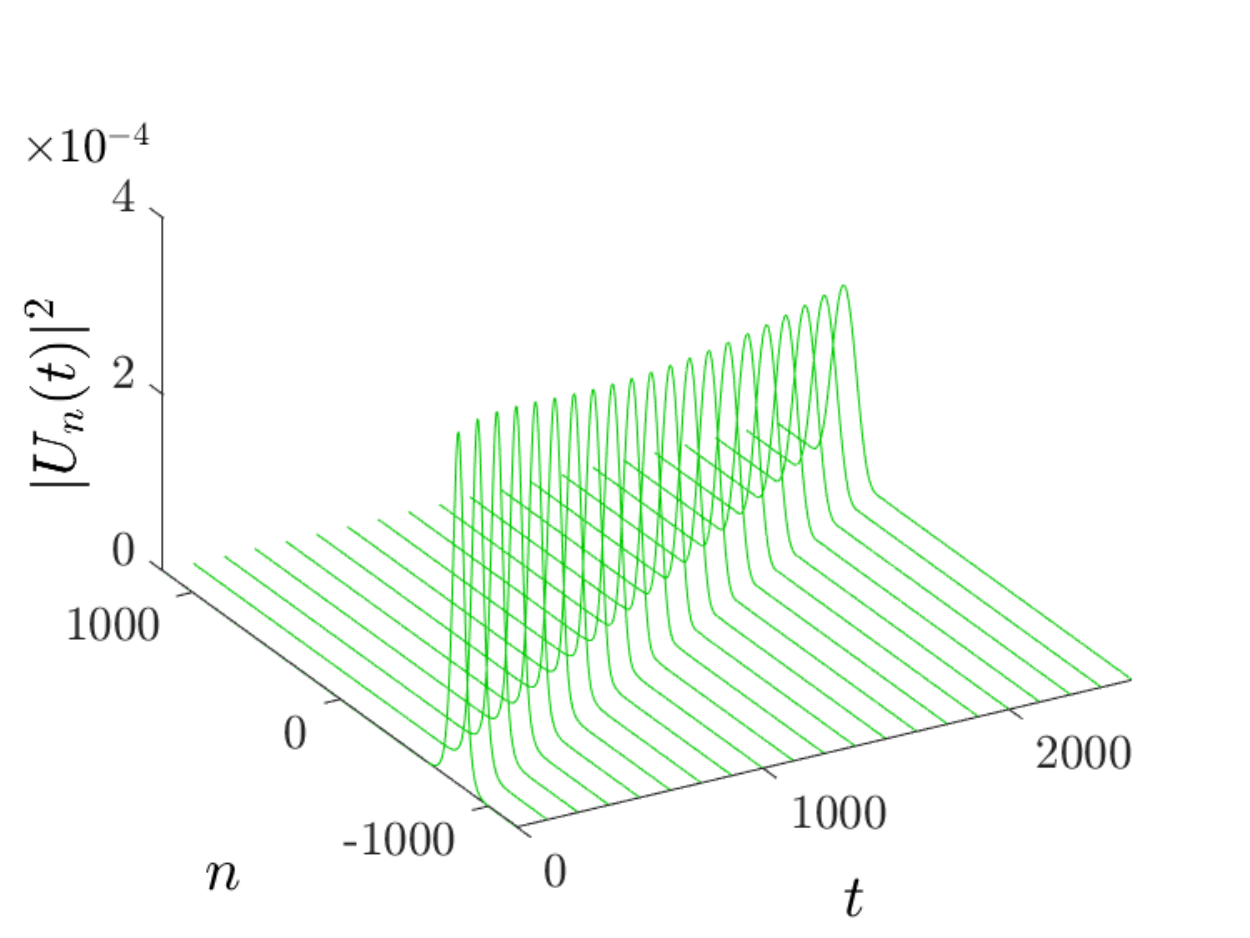}			
			\includegraphics[width=.50\textwidth]{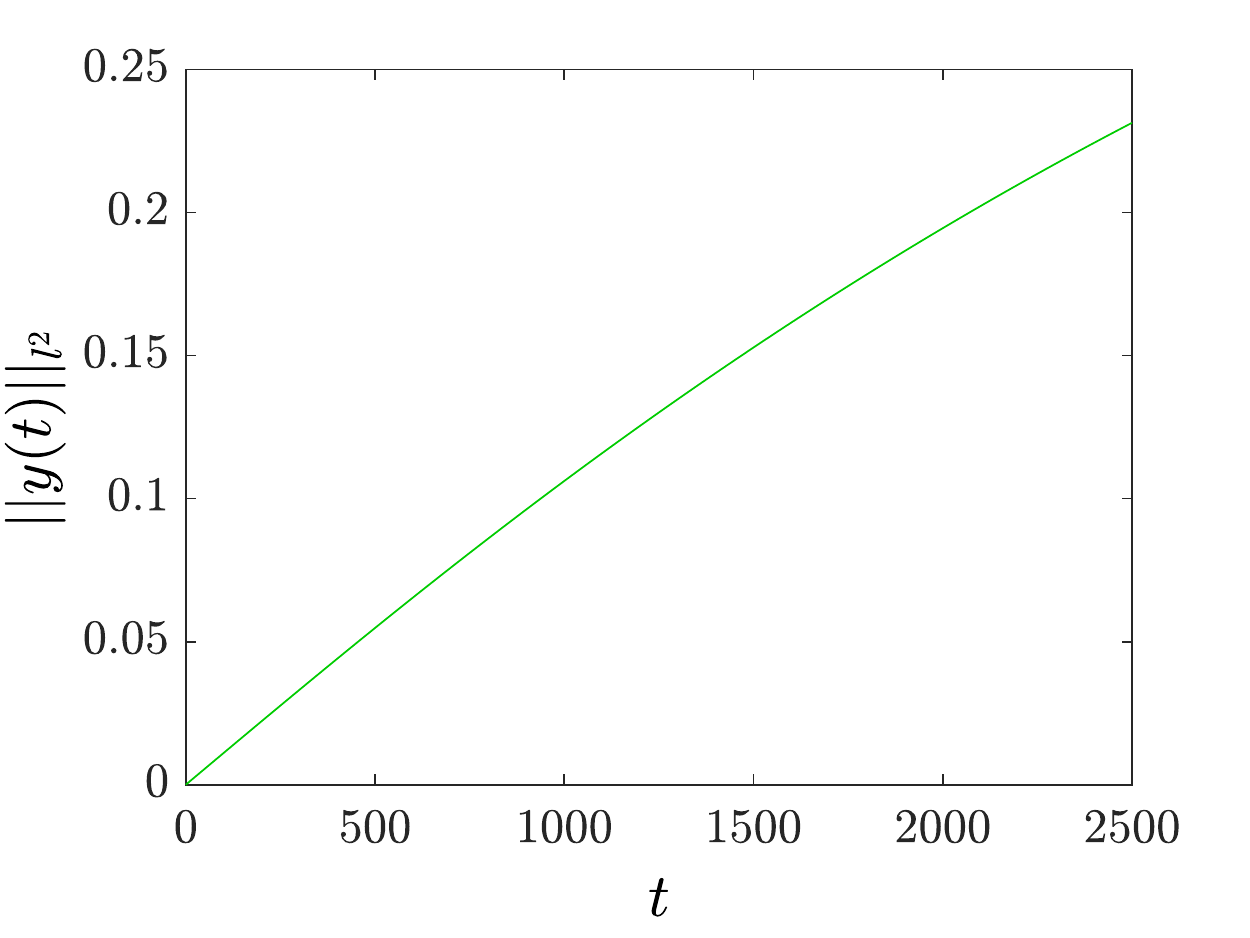}\\
			\includegraphics[width=.50\textwidth]{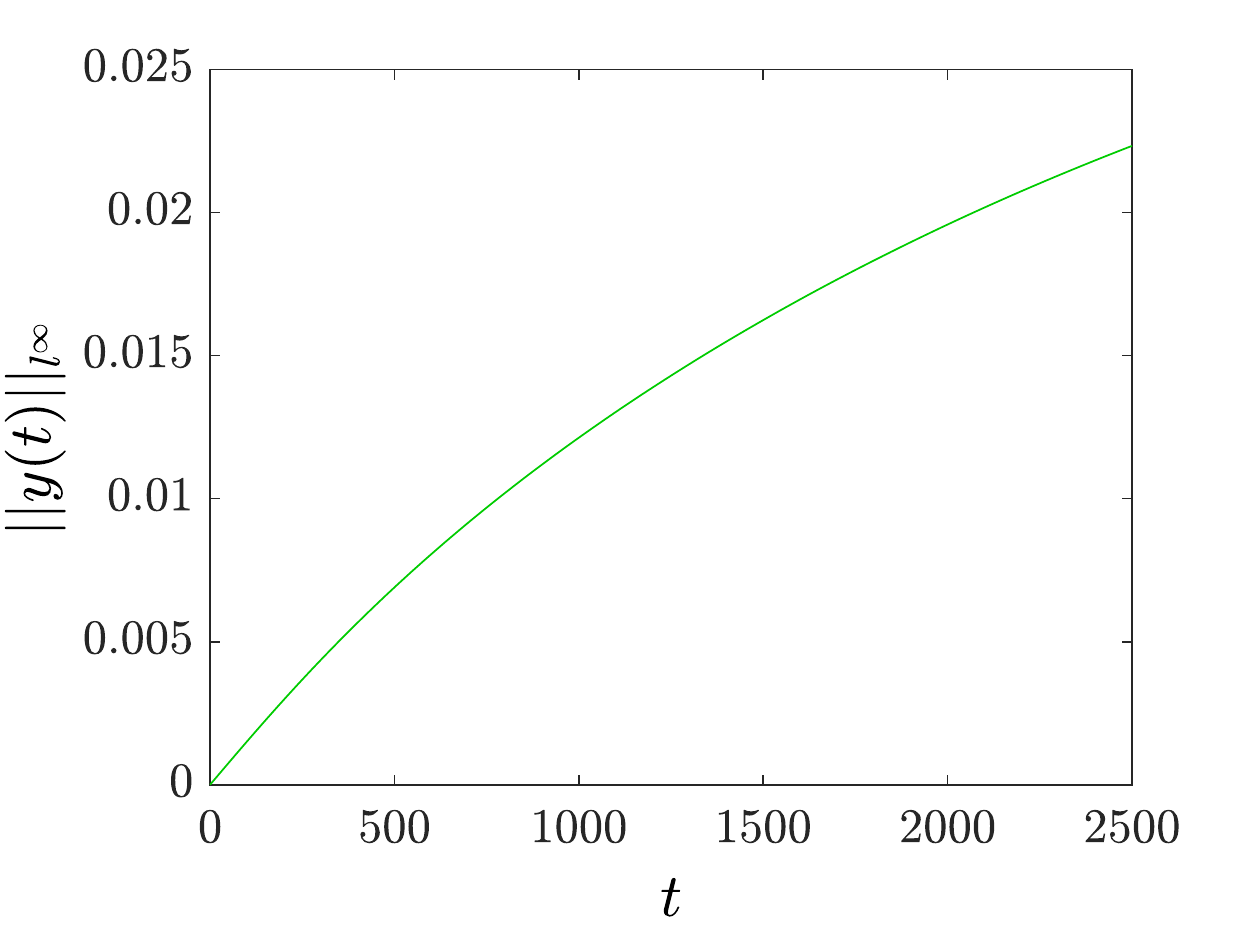}
			\includegraphics[width=.50\textwidth]{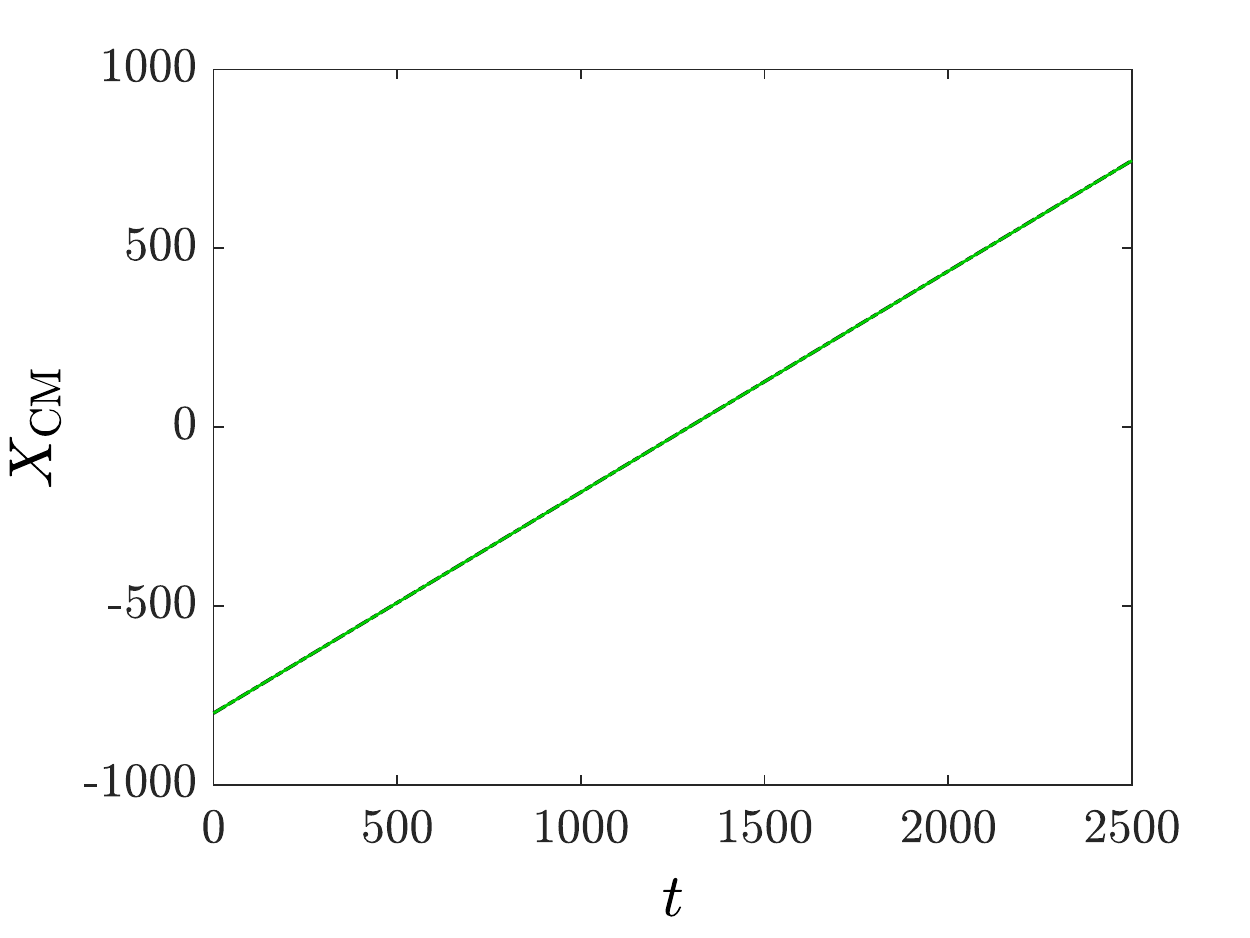}
		\end{tabular}
	\end{center}
	\caption{Top-left panel: Spatiotemporal evolution of the initial condition \eqref{sin1} in the G-AL lattice with $\delta=1$ for the quintic non-linearity $\sigma=0.2$. Parameters of the initial condition $\alpha=\pi/10$ and $\beta=\arcsinh(0.02)$, $\mu=1$. Top right-panel and bottom-left panel: Time evolution of $||y(t)||_{l^2}$ and $||y(t)||_{l^{\infty}}$, respectively  corresponding to the soliton dynamics shown in the top-left panel (details in the text of section \ref{secNum}). Bottom-right panel: Space time evolution of the soliton center $X_{\mathrm{CM}}=(\sum_n n|\phi_n|^2)/(\sum_n |\phi_n|^2)$  for the AL \eqref{eq:AL0} (black dash-dotted line) and G-AL \eqref{eq:AL0s} (green continuous line) .
	}
	\label{fig1}
\end{figure}
\begin{figure}[tbp!]
	\begin{center}
		\begin{tabular}{cc}
			\includegraphics[width=.50\textwidth]{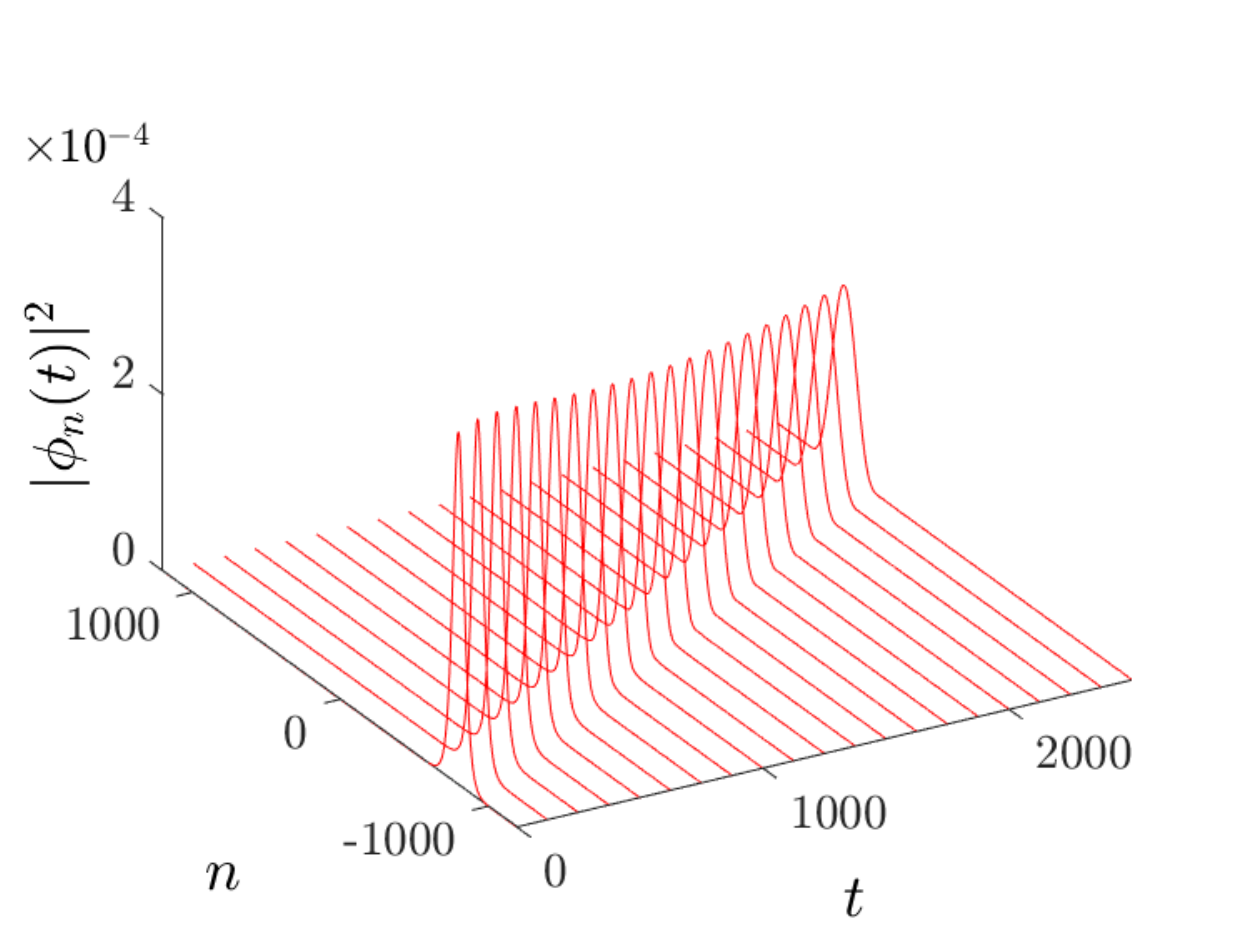}			
			\includegraphics[width=.50\textwidth]{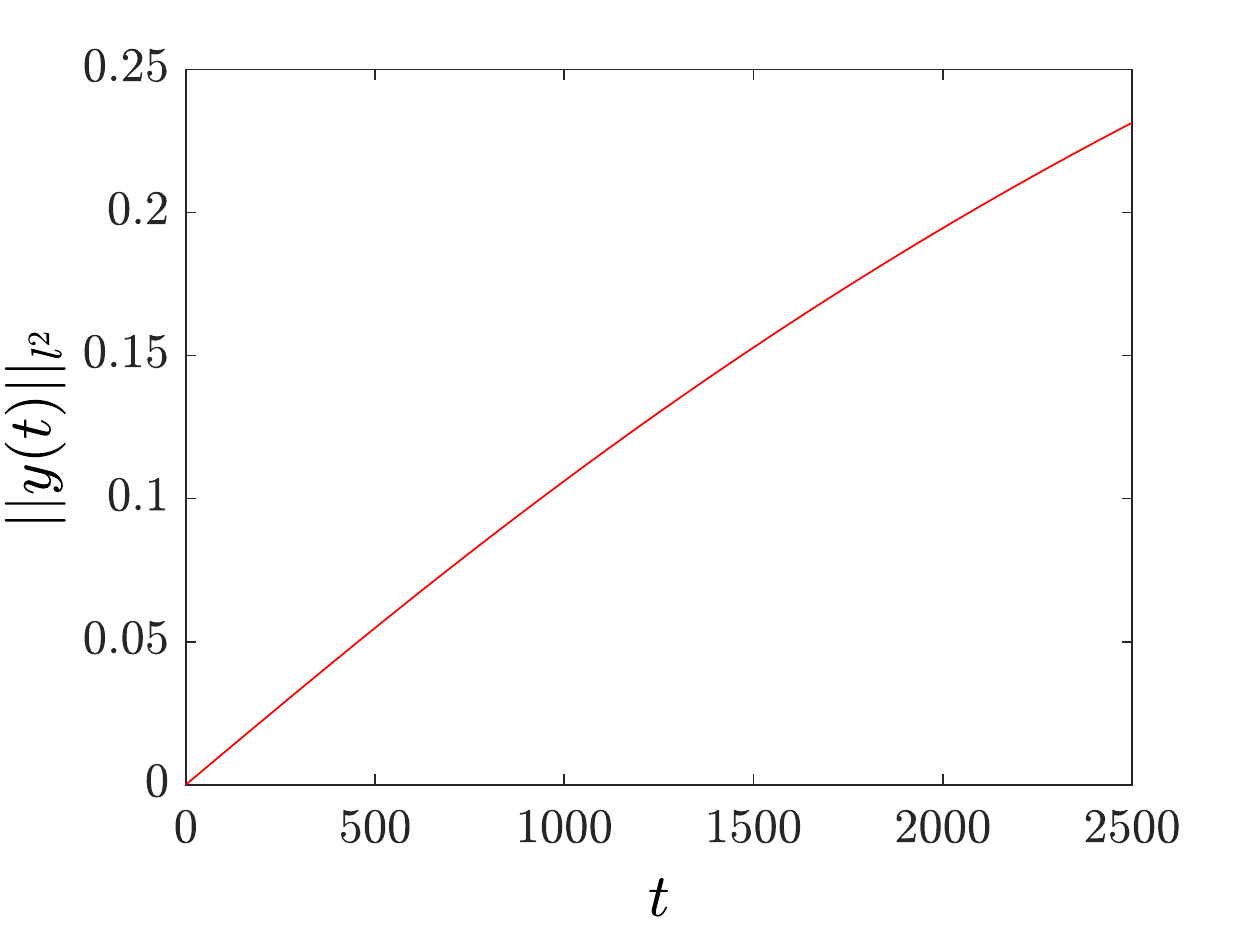}\\
			\includegraphics[width=.50\textwidth]{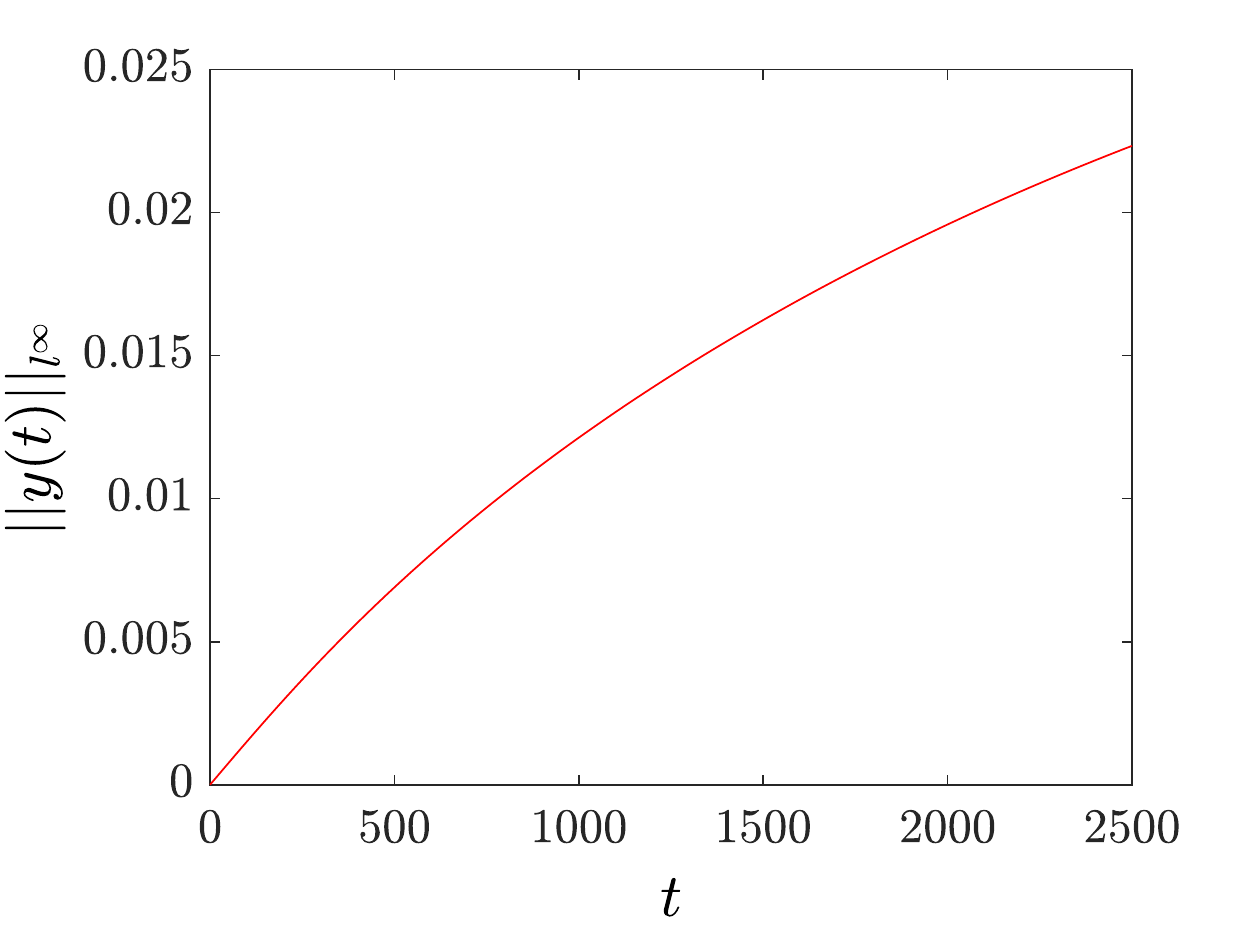}
			\includegraphics[width=.50\textwidth]{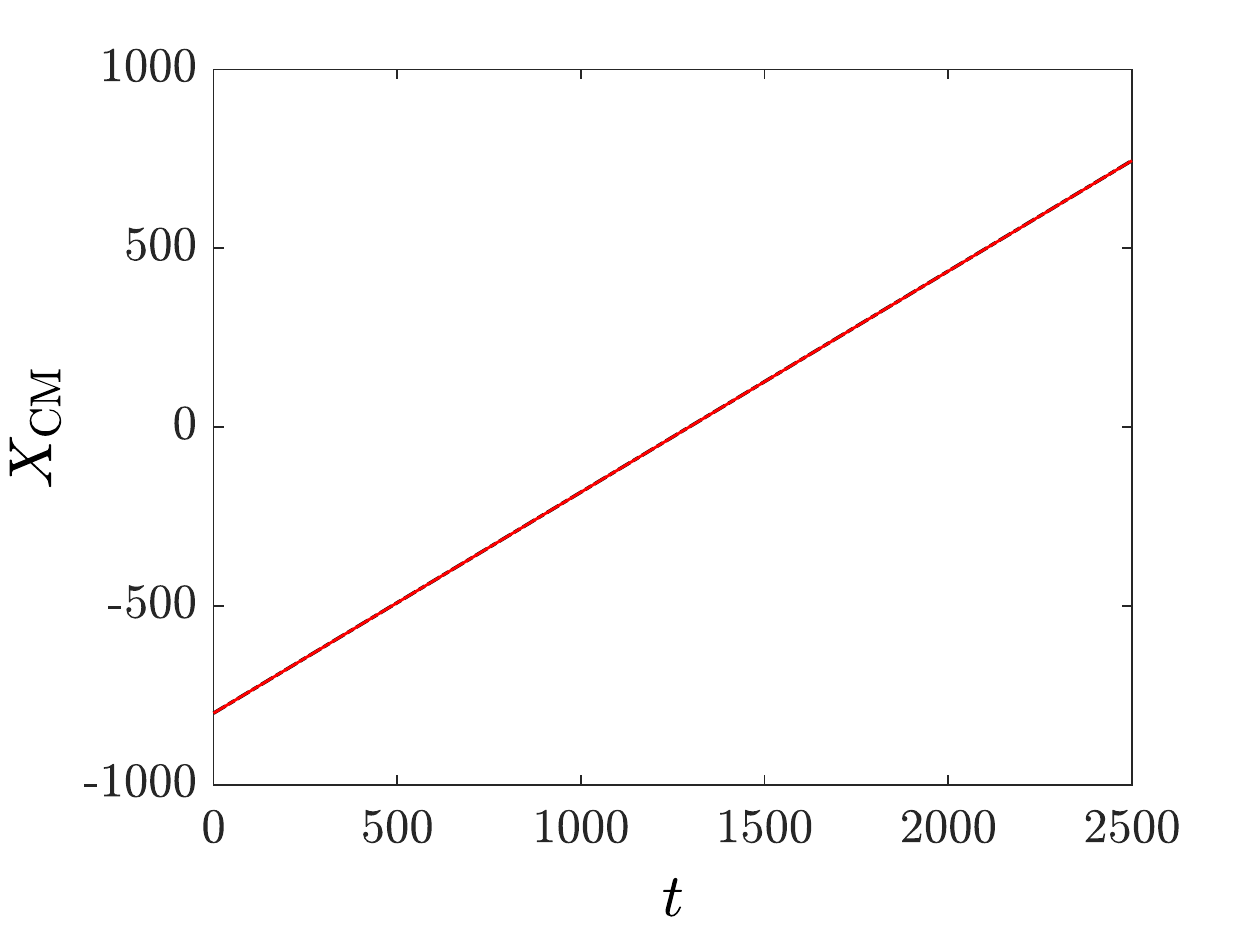}
		\end{tabular}
	\end{center}
	\caption{Top-left panel: Spatiotemporal evolution of the initial condition \eqref{sin1} in the quintic DNLS lattice  \eqref{eq:DNLS0s} with $\sigma=2$ and $\gamma=1$. Parameters of the initial condition $\alpha=\pi/10$ and $\beta=\arcsinh(0.02)$, $\mu=1$. Top right-panel and bottom-left panel: Time evolution of $||y(t)||_{l^2}$ and $||y(t)||_{l^{\infty}}$, respectively  corresponding to the soliton dynamics shown in the top-left panel (details in the text of section \ref{secNum}). Bottom-right panel: Space time evolution of the soliton center $X_{\mathrm{CM}}=(\sum_n n|\phi_n|^2)/(\sum_n |\phi_n|^2)$  for the AL \eqref{eq:AL0} (black dash-dotted line) and the quintic DNLS \eqref{eq:DNLS0s} (red continuous line) .
	}
	\label{fig2}
\end{figure}
\begin{figure}[tbp!]
	\begin{center}
		\begin{tabular}{cc}
			\includegraphics[width=.49\textwidth]{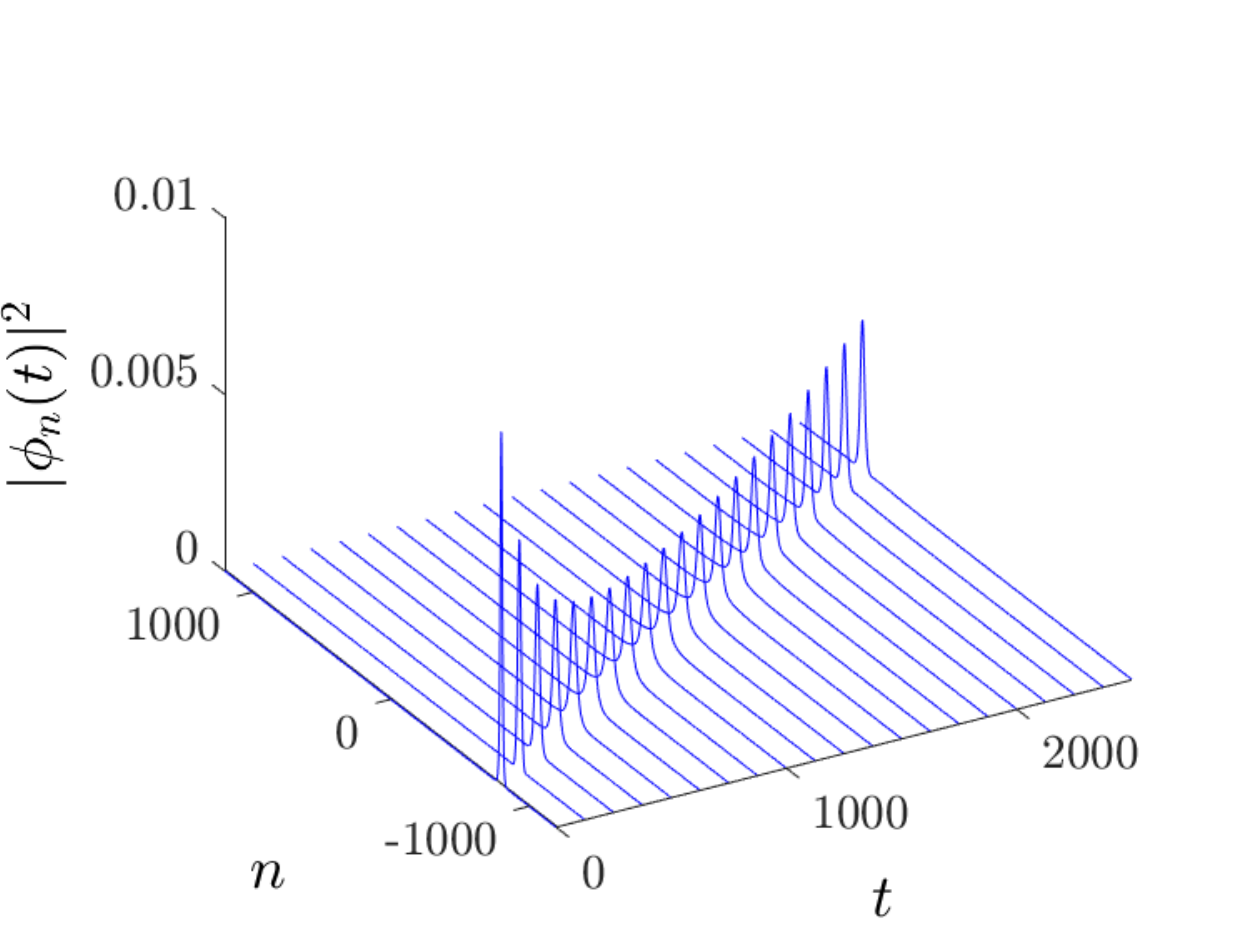}			
			\includegraphics[width=.49\textwidth]{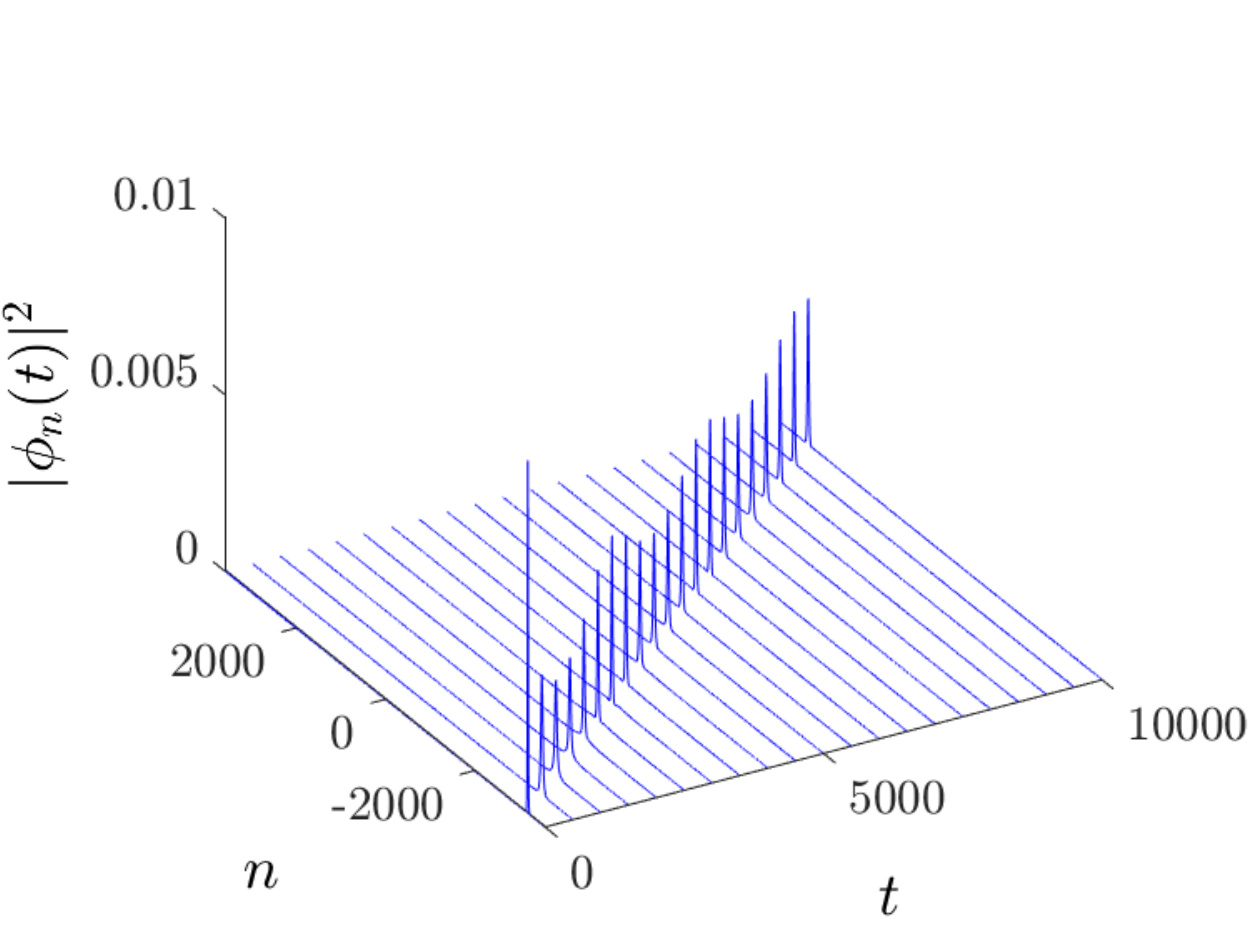}\\
			\includegraphics[width=.49\textwidth]{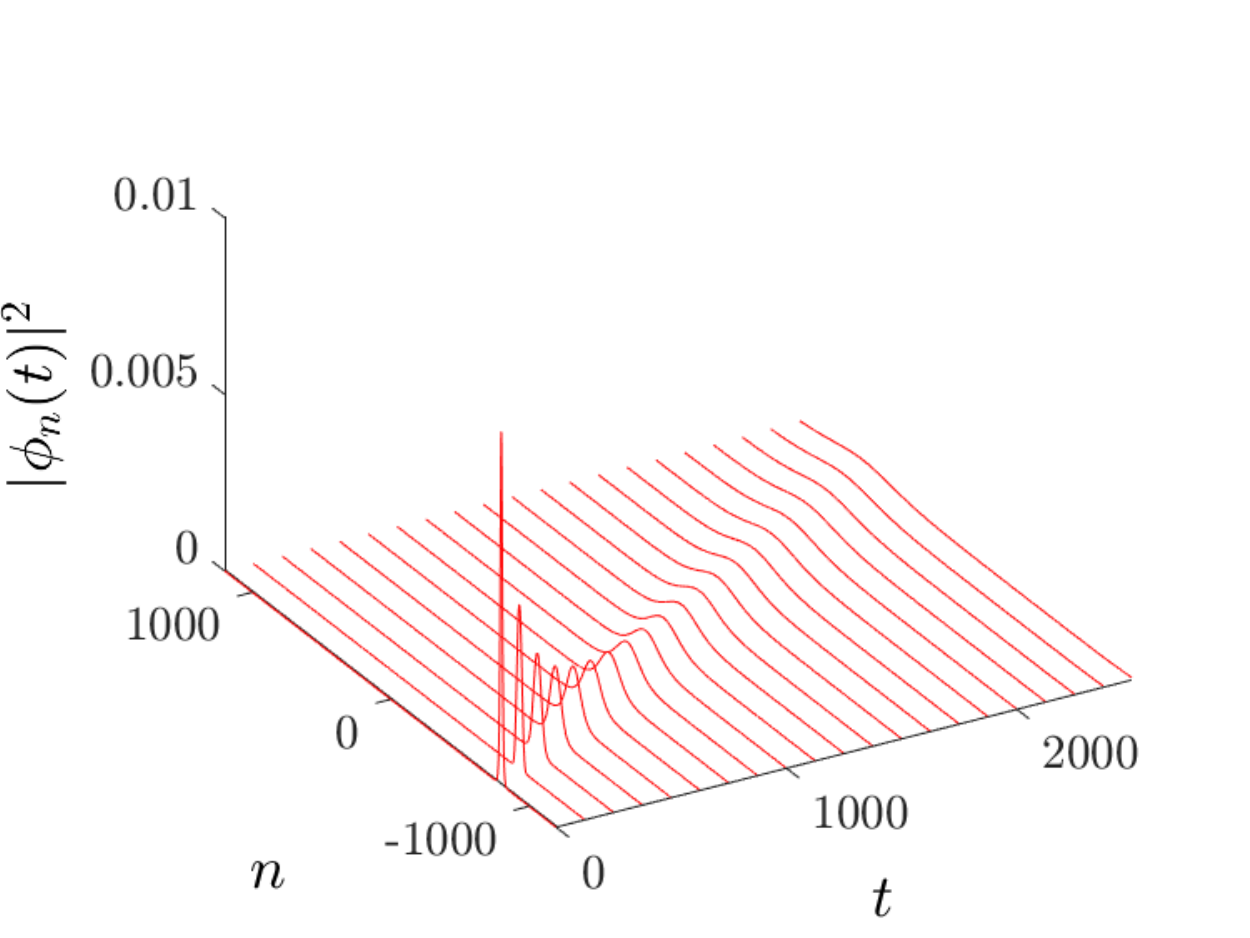}
			\includegraphics[width=.49\textwidth]{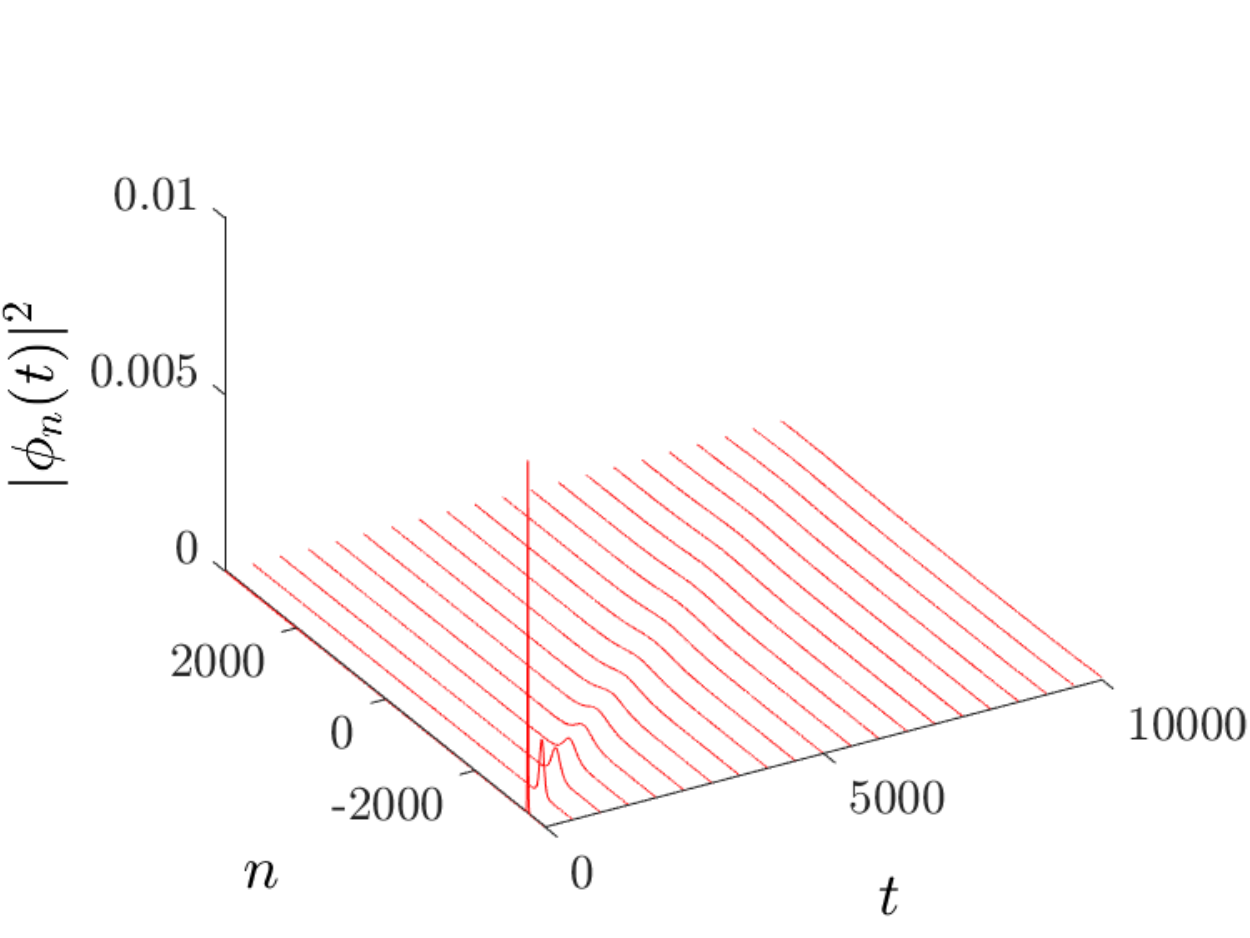}
		\end{tabular}
	\end{center}
	\caption{Top-left panel: Spatiotemporal evolution of the initial condition \eqref{sin1} in the cubic DNLS lattice  \eqref{eq:DNLS0s} with $\sigma=1$ and $\gamma=1$, for $t\in [0,T_f]=[0,2500]$, in a lattice of $K=2400$ units. Parameters of the initial condition $\alpha=\pi/10$ and $\beta=\arcsinh(1)$, $\mu=1$. Top right-panel: The same evolution for $t\in [0,T_f]=[0,10000]$ and for a chain of $K=7200$ units.  Bottom panels: The spatiotemporal evolution of the of the initial condition \eqref{sin1} with $\alpha=\pi/10$ and $\beta=\arcsinh(1)$, $\mu=1$ in the case of the quintic DNLS lattice \eqref{eq:DNLS0s}, $\sigma=2$. Rest of parameters as in the bottom panels.}
	\label{fig3}
\end{figure}
\begin{figure}[tbp!]
	\begin{center}
		\begin{tabular}{cc}
			\includegraphics[width=.49\textwidth]{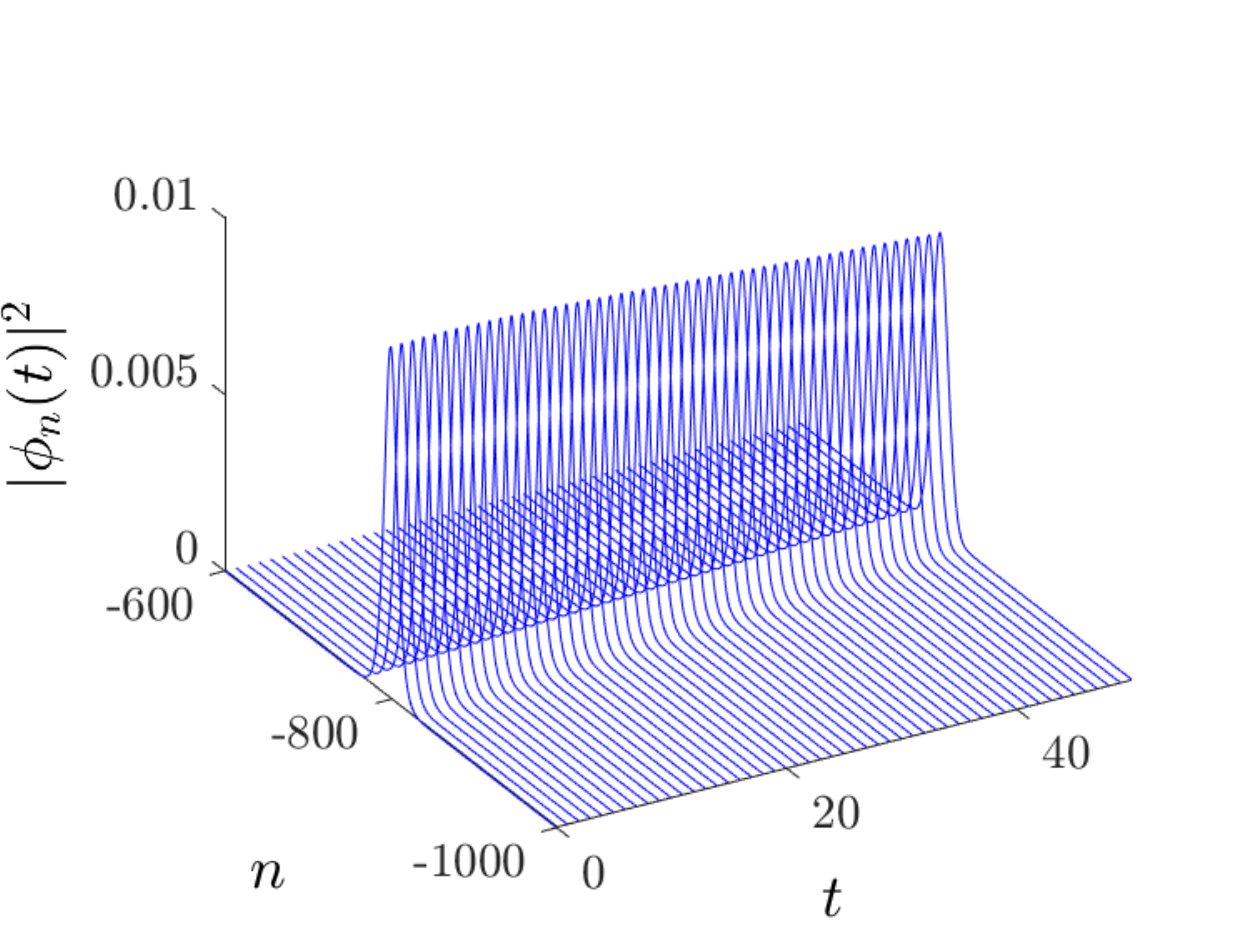}			
			\includegraphics[width=.49\textwidth]{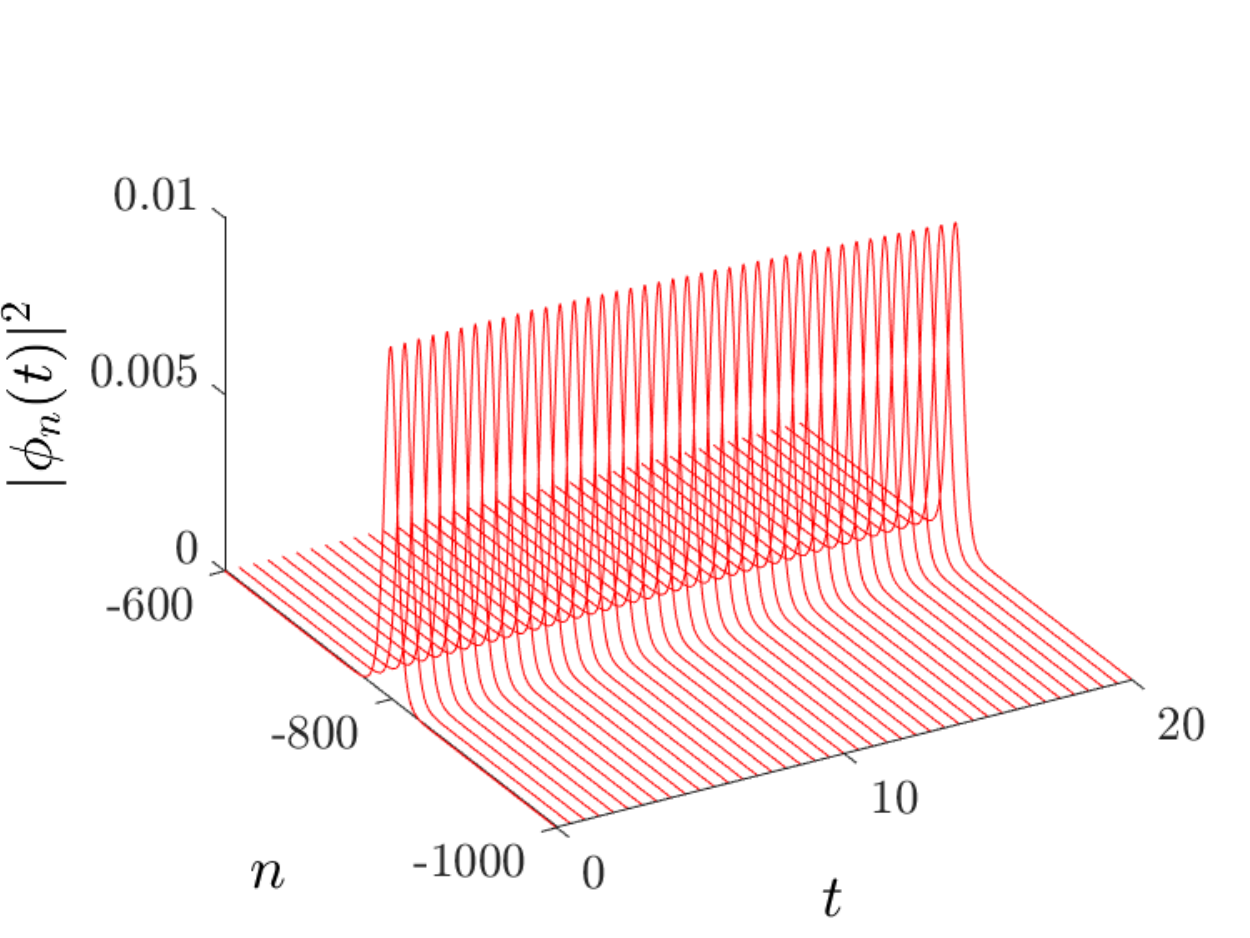}
		\end{tabular}
	\end{center}
	\caption{Same as figure \ref{fig3} but focusing on the time-interval $t\in [0,T_f]=[0,50]$ for $K=2400$ units.}
	\label{fig3b}
\end{figure}
\begin{figure}[tbp!]
	\begin{center}
		\begin{tabular}{cc}
			\includegraphics[width=.49\textwidth]{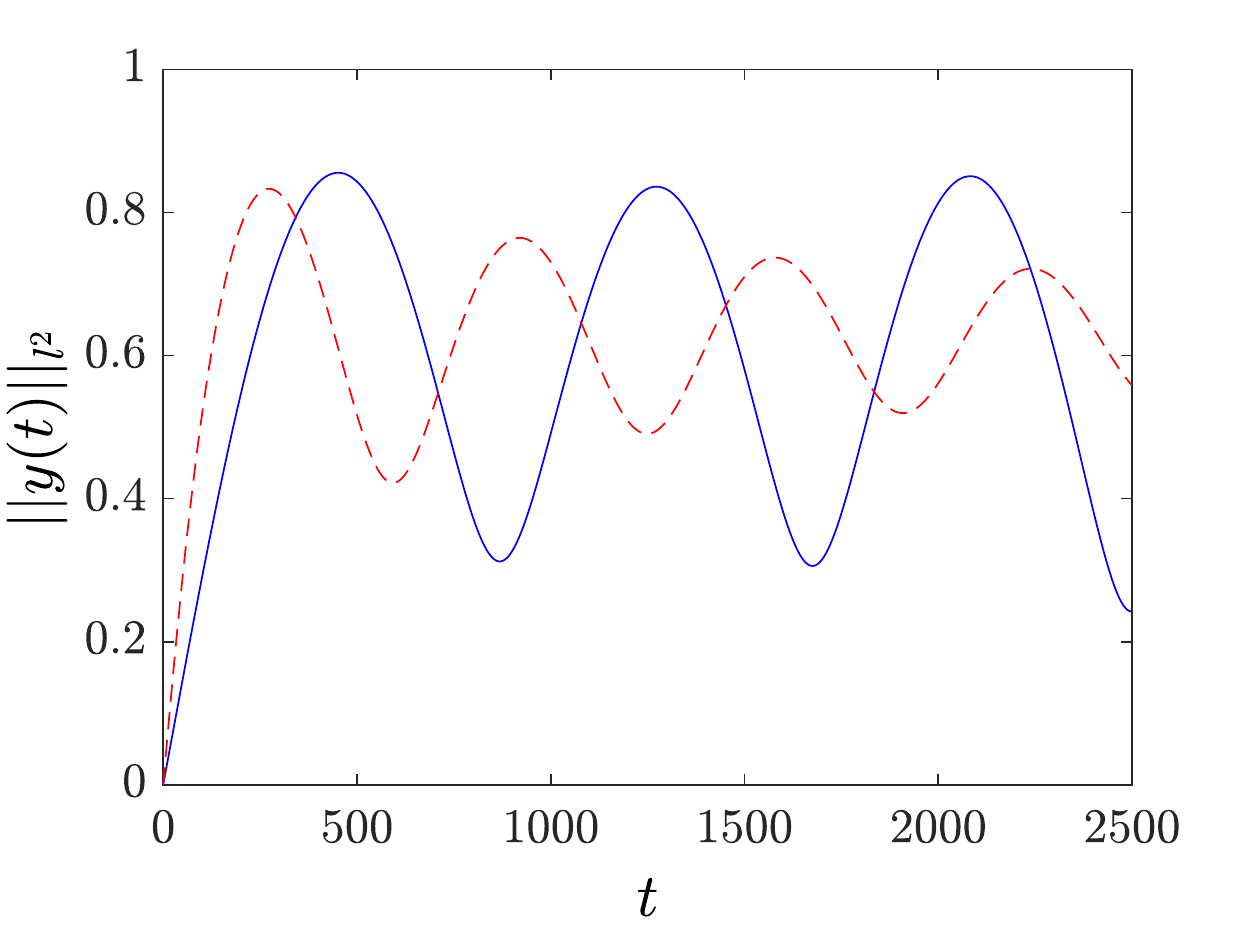}			
			\includegraphics[width=.49\textwidth]{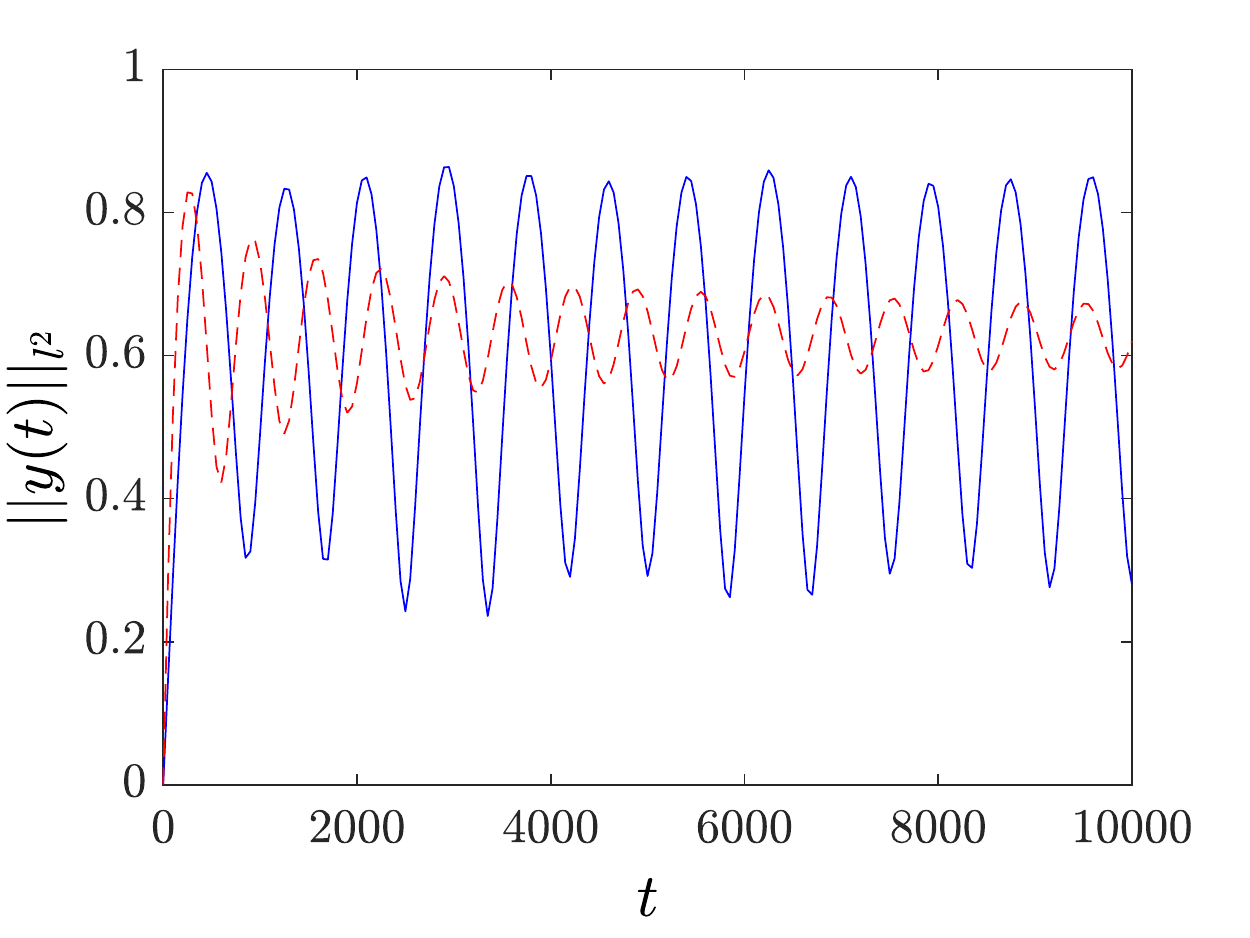}\\
			\includegraphics[width=.49\textwidth]{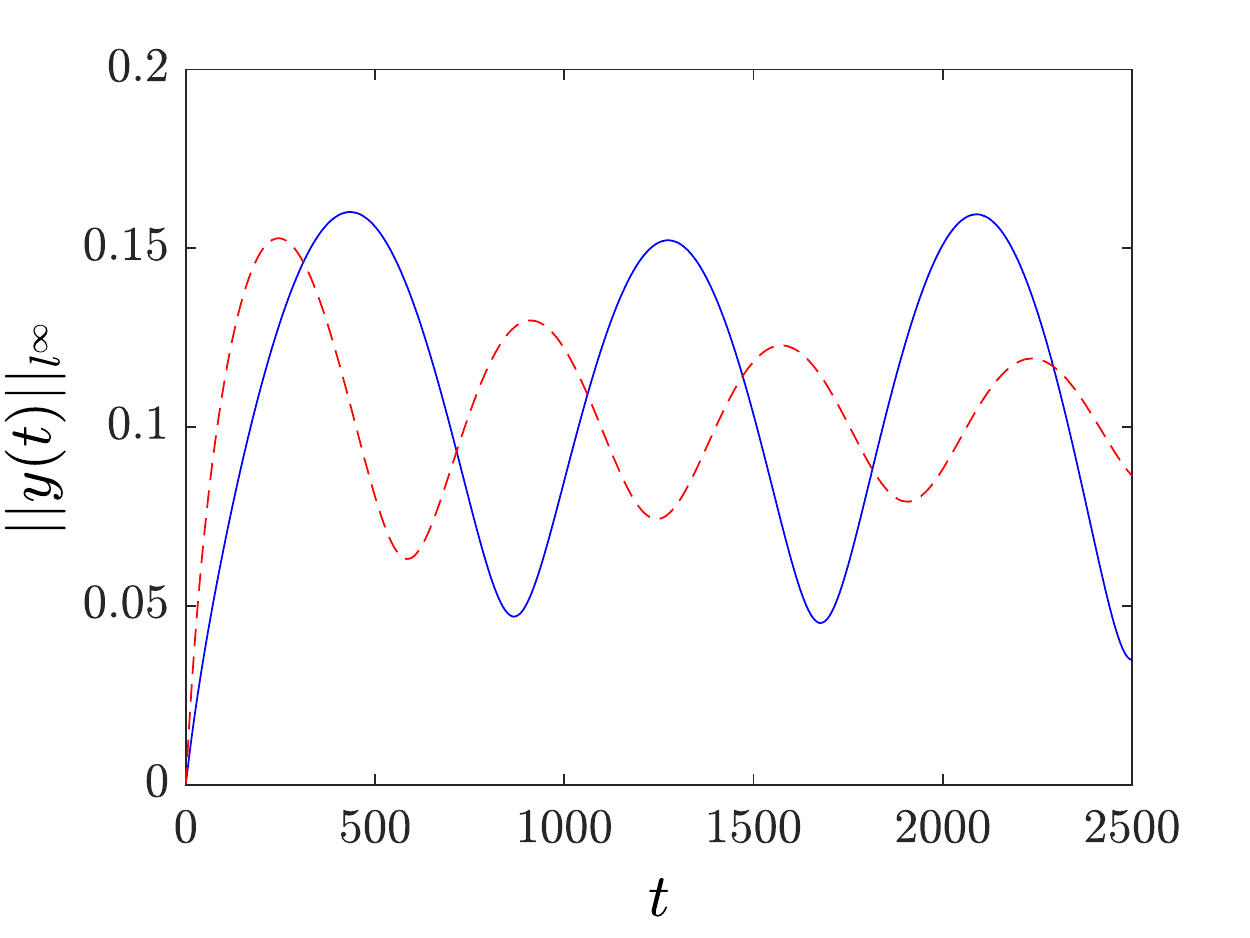}
			\includegraphics[width=.49\textwidth]{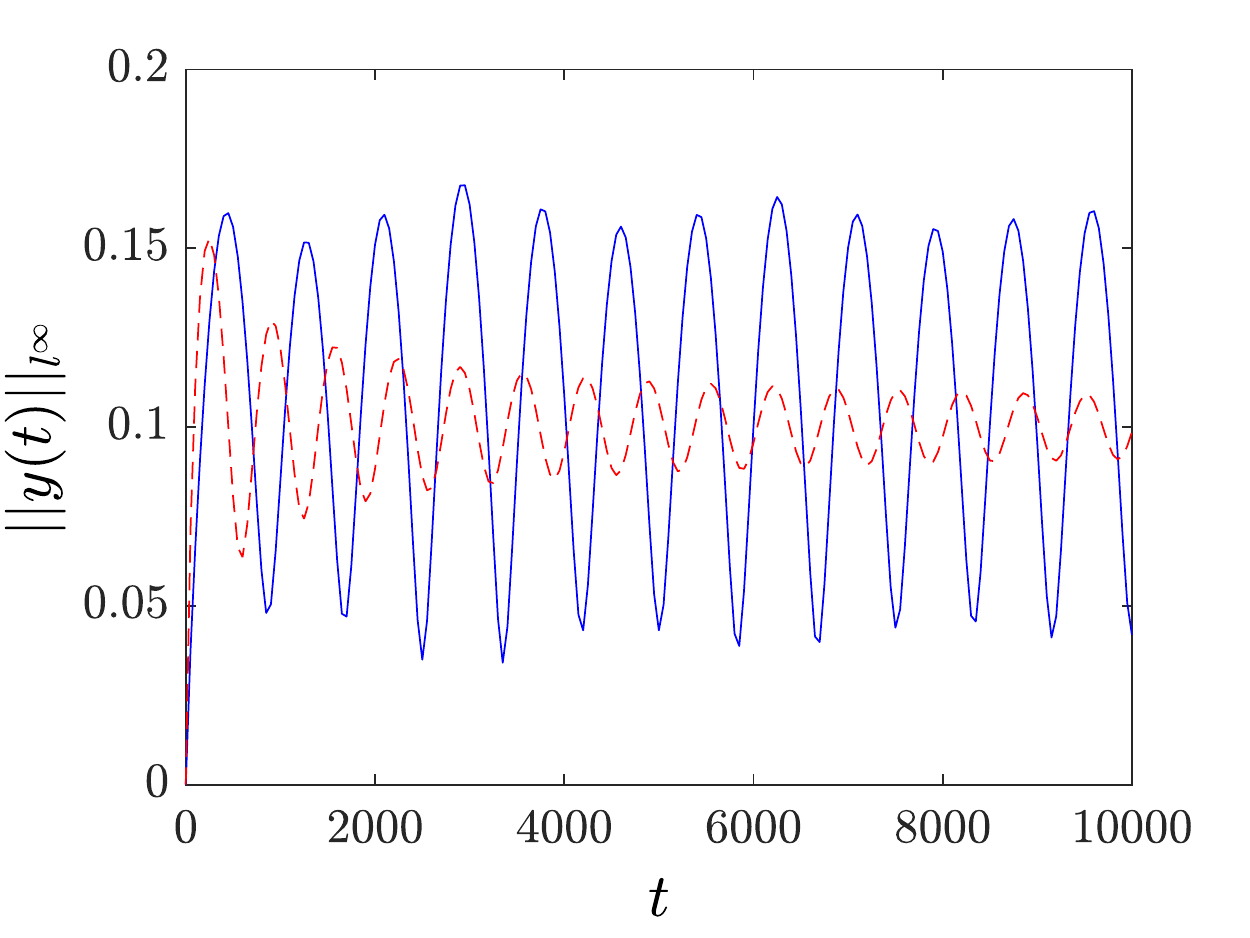}
		\end{tabular}
	\end{center}
	\caption{Top-left panel: Evolution of $||y(t)||_{l^2}$ associated to the spatiotemporal evolution in the cubic DNLS \eqref{eq:DNLS0s} $\sigma=1$-continuous blue curve and the quintic \eqref{eq:DNLS0s}  $\sigma=2$-dashed red curve when $t\in [0, 2500]$ and $K=2400$ lattice units. Top-right panel: Evolution of $||y(t)||_{l^2}$ associated to the spatiotemporal evolution in the cubic DNLS \eqref{eq:DNLS0s} $\sigma=1$-continuous blue curve and the quintic \eqref{eq:DNLS0s}  $\sigma=2$-dashed red curve when $t\in [0, 10000]$ and $K=7200$ lattice units. Bottom panels: Same as above but for  $||y(t)||_{l^{\infty}}$. }
	\label{fig4}
\end{figure}
\begin{figure}[tbp!]
	\begin{center}
		\begin{tabular}{cc}
			\includegraphics[width=.49\textwidth]{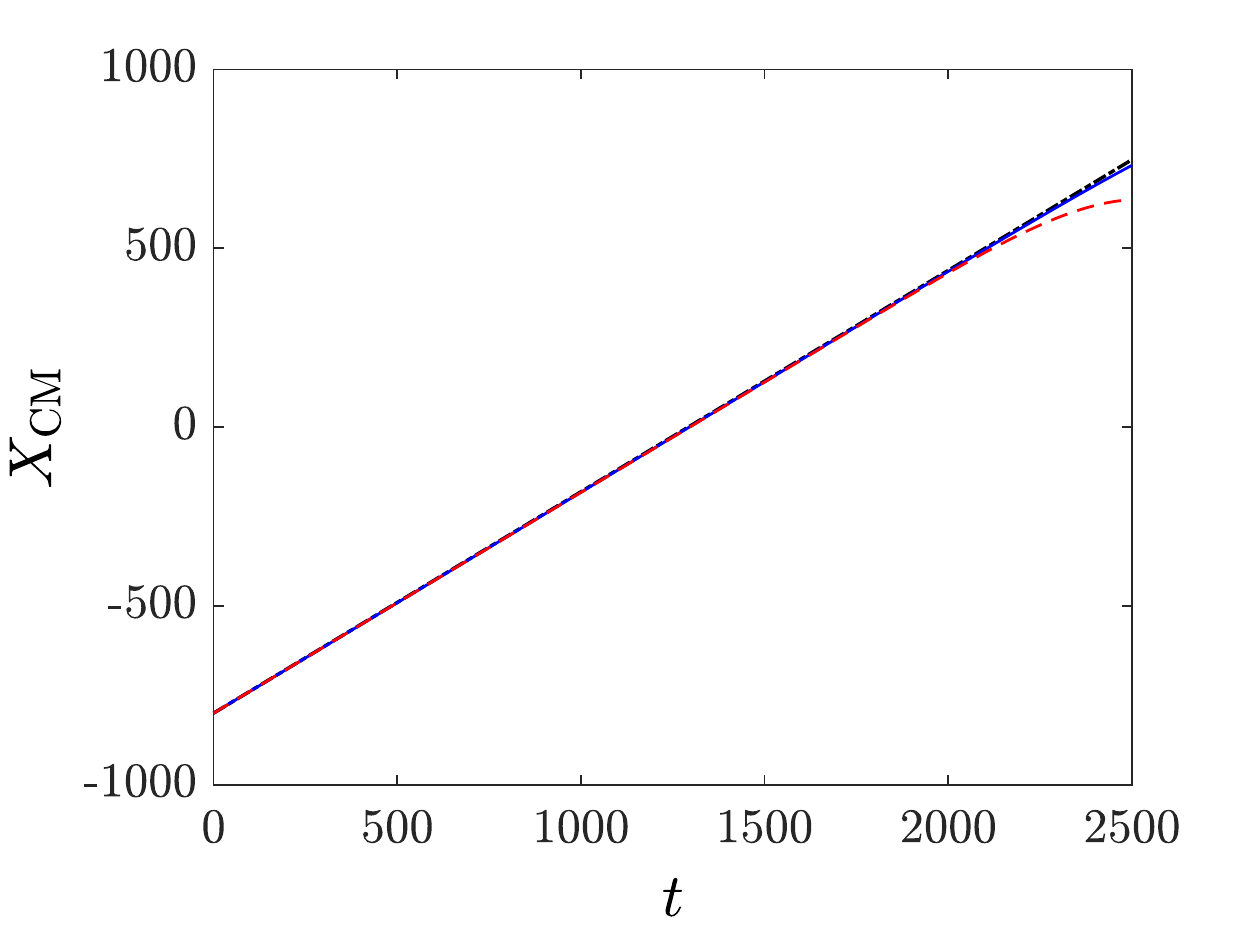}			
			\includegraphics[width=.49\textwidth]{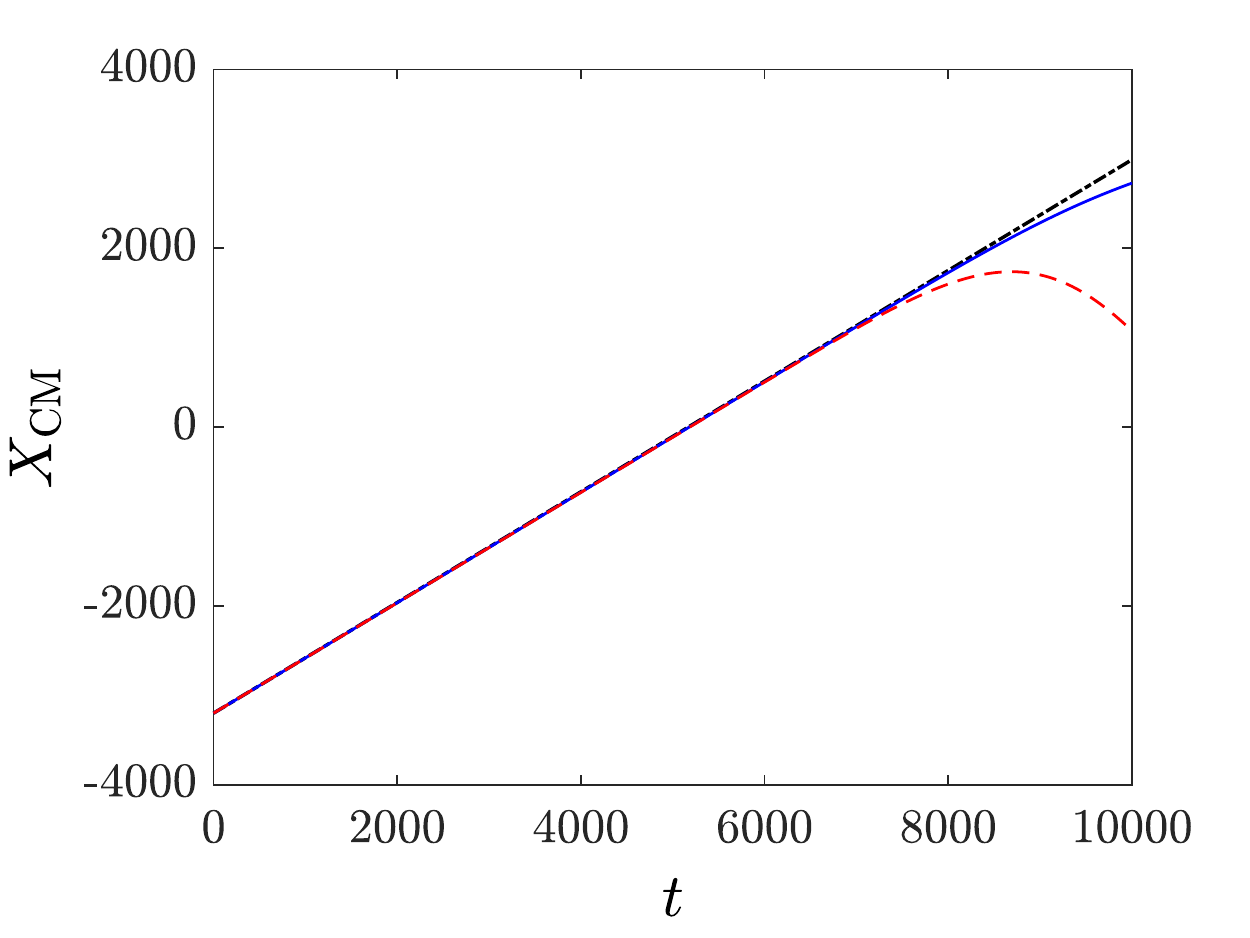}
		\end{tabular}
	\end{center}
	\caption{Space time evolution of the soliton center $X_{\mathrm{CM}}=(\sum_n n|\phi_n|^2)/(\sum_n |\phi_n|^2)$  for the AL \eqref{eq:AL0} (dashed-dotted black curve)  the cubic DNLS \eqref{eq:DNLS0s} $\sigma=1$ (continuous blue curve) and the quintic DNLS \eqref{eq:DNLS0s} $\sigma=2$ (continuous red curve), corresponding to the spatiotemporal evolution of the AL-soliton initial condition of Figure \ref{fig4} ($\beta=\arcsinh{0.1}$).  Left panel: For the lattices of $K=2500$ units and $t\in [0,2500]$. Right panel: For the lattices of $K=7200$ units and $t\in [0,10000]$. }
	\label{fig5}
\end{figure}
\section{Proof of closeness of the AL and G-AL solutions}
\label{SecII}
The analytical proof of closeness we shall demonstrate concerns the infinite lattice with vanishing boundary conditions. However we remark that the argument can be easily modified to cover other cases of boundary conditions such as periodic ones.  Consider then, the initial conditions for the integrable AL \eqref{eq:AL0} and the non-integrable G-AL
\begin{equation}
	\label{inAL0}
	\psi_n(0)=\phi_{n,0},\,\,\,n\in {\mathbb{Z}},
\end{equation}
and
\begin{equation}
	\label{ingAL}
	U_n(0)=U_{n,0},\,\,\,n\in {\mathbb{Z}},
\end{equation}
respectively.  We start with the statement of the following theorem.
\begin{theorem}
	\label{Theorem:closeness}
	a. We assume that  for every $0<\epsilon<1$,
	the initial conditions \eqref{inAL0} of the AL \eqref{eq:AL0}  and the initial conditions \eqref{ingAL} of the G-AL \eqref{eq:AL0s} satisfy:
	\begin{eqnarray}
		&&|| \psi(0)-U(0)||_{l^2}\le C_0\, \varepsilon^3,\label{con1}\\
		&&|| U(0)||_{l^2}\le C_{\delta,0}\, \varepsilon,\label{con2}\\
		&&P_{\mu}(0)=\sum_{n}\ln\left(1+\mu|\psi_n(0)|^2\right)\nonumber\\\
		&&\le \ln\left(1+(C_{\mu,0}\,\varepsilon)^2\right),\label{con3}
	\end{eqnarray}
	for some constants $C_0, C_{\delta,0},\,C_{\mu,0}>0$.	
	Then, for arbitrary $T_{\tiny{f}}>0$, there exists a constant $C=C(\delta,\mu,C_{\mu,0},C_{\delta,0},T_{\tiny{f}},\sigma)$, such that the corresponding solutions
	for every $t\in [0,T_{\tiny{f}}]$, satisfy the estimate
	\begin{equation}
		||y(t)||_{l^2}=|| \psi(t)-U(t)||_{l^2}\le C \varepsilon^3.\label{eq:boundy}
	\end{equation}
	b. When $\phi(0)=U(0)$, under the assumptions \eqref{con2}-\eqref{con3}, for every $\varepsilon>0$ and $t\in [0,T_{\tiny{f}}]$, the maximal distance $|| y(t)||_{l^\infty}=\sup_{n\in {\mathbb{Z}}}|y_n(t)|=\sup_{n\in {\mathbb{Z}}}|\psi_n(t)-U_n(t)|$ between individual units of the systems satisfies the estimate
	\begin{equation}
		||y(t)||_{l^\infty}\le C \varepsilon^3.\label{eq:boundyNN}
	\end{equation}
\end{theorem}
\textbf{Proof:} For simplicity, we set $\kappa=\nu=1$. The corresponding evolution equation for the local distance $y_n=\psi_n-U_n$ reads
\begin{equation}
	\label{loc1}
	i\dot{y}_n=\Delta_dy_n+\left[\mu|\psi_n|^{2}(\psi_{n+1}+\psi_{n-1})-\delta\,|U_{n}|^{2\sigma}(U_{n+1}+U_{n-1})\right],
\end{equation}
where $\Delta_dy_n=y_{n+1}-2y_n+y_{n-1}$ is the 1D-discrete Laplacian. Let us recall that when $u_n\in\mathbb{C}$, the space $l^2(\mathbb{Z};\mathbb{C})$ becomes a real Hilbert space, $l^2(\mathbb{Z};\mathbb{C})\equiv l^2(\mathbb{Z};\mathbb{R})\times l^2(\mathbb{Z}^N;\mathbb{R})$, if endowed with the real inner product
\begin{eqnarray*}
	(u,v)_{l^2}=\mathrm{Re}\sum_{n\in\mathbb{Z}^N}u_n\bar{v}_n.
\end{eqnarray*}
In this setting the operator $\Delta_d:l^2\rightarrow l^2$ is bounded and self-adjoint,
\begin{equation*}
	\begin{split}
		(\Delta_{d}u,v)_{l^2}&=(u,\Delta_{d}v)_{l^2}, \quad u,\;v\in l^2,\\
		||\Delta_d u||_{l^2}^2&\leq ||u||_{l^2}^2.
	\end{split}
\end{equation*}
Moreover, it satisfies
\begin{equation}
	\label{loc2}
	(\Delta_{d}u,u)_{l^2}=-\sum_{n\in\mathbb{Z}}|u_{n+1}-u_n|^2\leq 0,\,\,\mbox{for all}\,\, u\in l^2.
\end{equation}
When multiplying Eq. \eqref{loc1} by $\overline{y}_n$, summing over $\mathbb{Z}$, and keeping the imaginary parts, we may use \eqref{loc3} and get that
\begin{equation}
	\label{loc3}
	\begin{split}
		\frac{1}{2}\frac{d}{dt}|| y(t)||_{l^2}^2=
		\mu\mathrm{Im}\sum_{n\in\mathbb{Z}^N}\left[|\psi_n|^{2}(\psi_{n+1}+\psi_{n-1})\right]\overline{y}_n-\delta\mathrm{Im}\sum_{n\in\mathbb{Z}^N}\left[|U_n|^{2\sigma}(U_{n+1}+U_{n-1})\right]\overline{y}_n
	\end{split}
\end{equation}
To estimate the terms on the right-hand side of \eqref{loc3} we  use the Cauchy-Schwarz inequality and the continuous embeddings
\begin{equation}
	\label{embe}
	l^r\subset l^s,\,\,\,|| \phi||_{l^s}\le || \phi||_{l^r},\,\,\,1 \le r\le s \le \infty.
\end{equation}
For the first term on the right-hand side of \eqref{loc3} we get:
\begin{equation}
	\label{loc5}
	\begin{split}
		\left|\mathrm{Im}\sum_{n\in\mathbb{Z}}\left[|\psi_n|^{2}(\psi_{n+1}+\psi_{n-1})\right]\overline{y}_n\right|
		&\leq 2\sup_{n\in\mathbb{Z}}|\psi_n|\sum_{n\in\mathbb{Z}^N}|\psi_n|^{2}\,|\overline{y}_n|\\
		&\leq 2\left(\sup_{n\in\mathbb{Z}}|\psi_n|\right)^{2}\sum_{n\in\mathbb{Z}^N}|\psi_n|\,|\overline{y}_n|
		\leq 2\left(\sup_{n\in\mathbb{Z}^N}|\psi_n|\right)^{2}||\phi||_{l^2}||y||_{l^2}\\
		&\leq 2||\psi||_{l^2}^{3}||y||_{l^2}.
	\end{split}
\end{equation}
For the second term on the right-hand side of \eqref{loc3} we obtain:
\begin{equation}
\label{loc5s}
\begin{split}
\left|\mathrm{Im}\sum_{n\in\mathbb{Z}}\left[|U_n|^{2\sigma}(U_{n+1}+U_{n-1})\right]\overline{y}_n\right|
&\leq 2\sup_{n\in\mathbb{Z}}|U_n|\sum_{n\in\mathbb{Z}^N}|U_n|^{2\sigma}\,|\overline{y}_n|\\
&\leq 2\left(\sup_{n\in\mathbb{Z}}|U_n|\right)^{2\sigma}\sum_{n\in\mathbb{Z}^N}|U_n|\,|\overline{y}_n|
\leq 2\left(\sup_{n\in\mathbb{Z}^N}|U_n|\right)^{2\sigma}||U||_{l^2}||y||_{l^2}\\
&\leq 2||U||_{l^2}^{2\sigma+1}||y||_{l^2}.
\end{split}
\end{equation}
Then, from \eqref{loc3}-\eqref{loc5s} we get the inequality
\begin{equation}
	\label{loc6}
	\frac{1}{2}\frac{d}{dt}|| y(t)||_{l^2}^2\leq 2\left(\mu||\psi||_{l^2}^{3}+\delta ||U||_{l^2}^{2\sigma+1}\right)||y(t)||_{l^2}.
\end{equation}
The inequality \eqref{loc6}, when combined with
$\frac{d}{dt}|| y(t)||_{l^2}^2=2|| y(t)||_{l^2} \frac{d}{dt}|| y(t)||_{l^2}$, implies that
\begin{equation}
	\label{loc7}
	\frac{d}{dt}|| y(t)||_{l^2}\leq 2\left(\gamma||\psi||_{l^2}^{3}+\delta ||U||_{l^2}^{2\sigma+1}\right).
\end{equation}
For the solutions of the AL the {\em deformed} norm
$$ P_{\mu}(t)=\sum_{n}\ln(1+\mu|\psi_n|^2)$$
is conserved, that is
\begin{equation}
	P_{\mu}(t)=P_{\mu}(0).
\end{equation}
This conservation implies the bound (see \cite[Lemma II.1]{DNJ2021} and \cite{Kim}),
\begin{equation}
	\label{loc9}
	|| \psi(t) ||_{l^2}^2\le \exp(P_{\mu}(0))-1\leq C^2_{0,\mu}\varepsilon^2,\qquad \forall t\ge 0.
\end{equation}
On the other hand, the G-AL, has a nontrivial conservation law \cite{JGAL}:
\begin{equation}
\label{loc10h}
P_\delta(t)=\sum_{n\in\mathbb{Z}}|U_n|^2\,_2F_1\bigg(1,\frac{1}{\sigma},1+\frac{1}{\sigma};-|U_n|^2\bigg),
\end{equation}
where $_2F_1(a,b,c;z)$ is the Gauss hypergeometric function, for $a,b,c,z\in\mathbb{C}$. We recall that
\begin{eqnarray}
\label{loc11h}
_2F_1(a,b,c;z)=1+\frac{ab}{c}z+\frac{a(a+1)b(b+1)}{c(c+1)2!}z^2+...,
\end{eqnarray}
on the disk $|z|<1$ and by analytic continuation outside this disk.  Under the assumption \eqref{con2} the expansion \eqref{loc11h} implies that
\begin{eqnarray*}
P_{\delta}(0)\leq C_{\delta,0}\epsilon^2+\mathcal{O}(\epsilon^4),
\end{eqnarray*}
and the conservation of \eqref{loc10h} shows that
\begin{eqnarray}
\label{loc12h}
||U(t)||_{l^2}^2\leq \hat{C}_{0,\sigma,\delta}\epsilon^2+\mathcal{O}(\epsilon^4),
\end{eqnarray}
where  $\hat{C}_{0,\sigma,\delta}$ will stand for a generic constant depending on $\sigma$ and $C_{\delta,0}$. Then, using \eqref{loc9} and \eqref{loc12h} in the differential inequality \eqref{loc7} we get that
\begin{equation}
	\label{loc10}
	\frac{d}{dt}|| y(t)||_{l^2}\leq 2\left(\mu C_{0,\mu}^3+\delta \hat{C}_{0, \delta,\sigma}\right)\varepsilon^3.
\end{equation}
Integrating the inequality \eqref{loc10} in the arbitrary interval $[0, T_{\tiny{f}}]$, and using the assumption \eqref{con1} on the distance $||y(0)||_{l^2}=||\phi(0)-y(0)||_{l^2}$ of the initial data, we obtain that
\begin{equation*}
	\begin{split}
		|| y(t)||_{l^2}&\leq 2\left(\mu C_{0,\mu}^3+\delta \hat{C}_{0,\sigma,\delta}\right)T_{\tiny{f}}\varepsilon^3+||y(0)||_{l^2}\\
		&\leq  2\left(\mu C_{0,\mu}^3+\delta \hat{C}_{0,\sigma,\delta}\right)T_{\tiny{f}}\varepsilon^3+C_0\varepsilon^3.
	\end{split}
\end{equation*}
Hence, for the constant
\begin{equation}
	\label{loc11}
	C=2\left(\mu C_{0,\mu}^3+\delta\hat{C}_{0,\sigma,\delta}\right))T_{\tiny{f}}+C_0,
\end{equation}
we conclude with the claimed estimate \eqref{eq:boundy}.

\textit{b.} The result is an immediate consequence of the embedding \eqref{embe} and the estimate \eqref{eq:boundy}, since $||y(t)||_{\l^{\infty}}\leq ||\phi(t)-\psi(t)||_{\l^2}$. \ \ $\square$
%
%
\section{Numerical Results}
\label{SecIII}
\setcounter{equation}{0}
\label{secNum}
%
%
\begin{figure}[tbp!]
	\begin{center}
		\begin{tabular}{cc}
			\includegraphics[width=.49\textwidth]{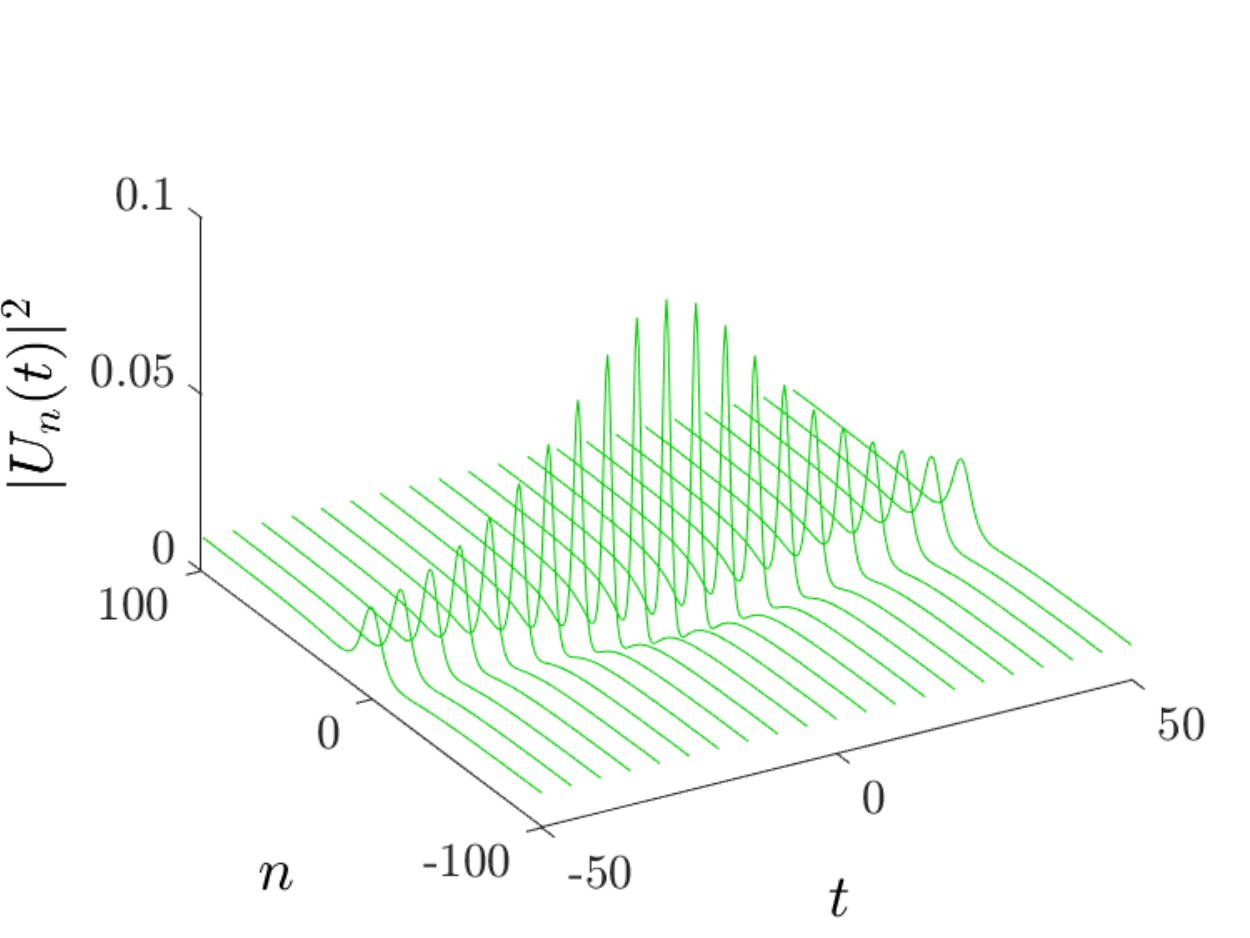}		
			\includegraphics[width=.49\textwidth]{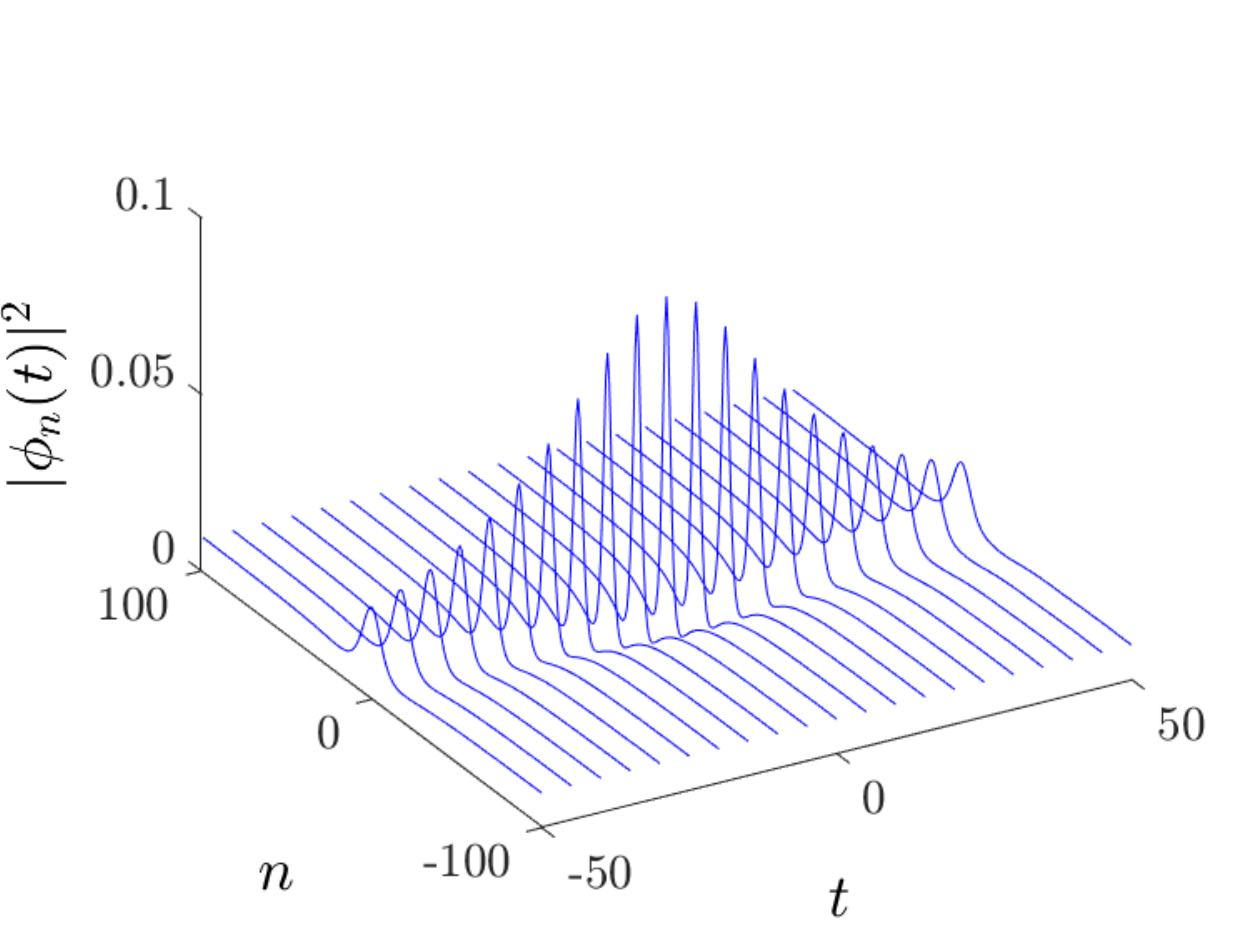}\\
			\includegraphics[width=.49\textwidth]{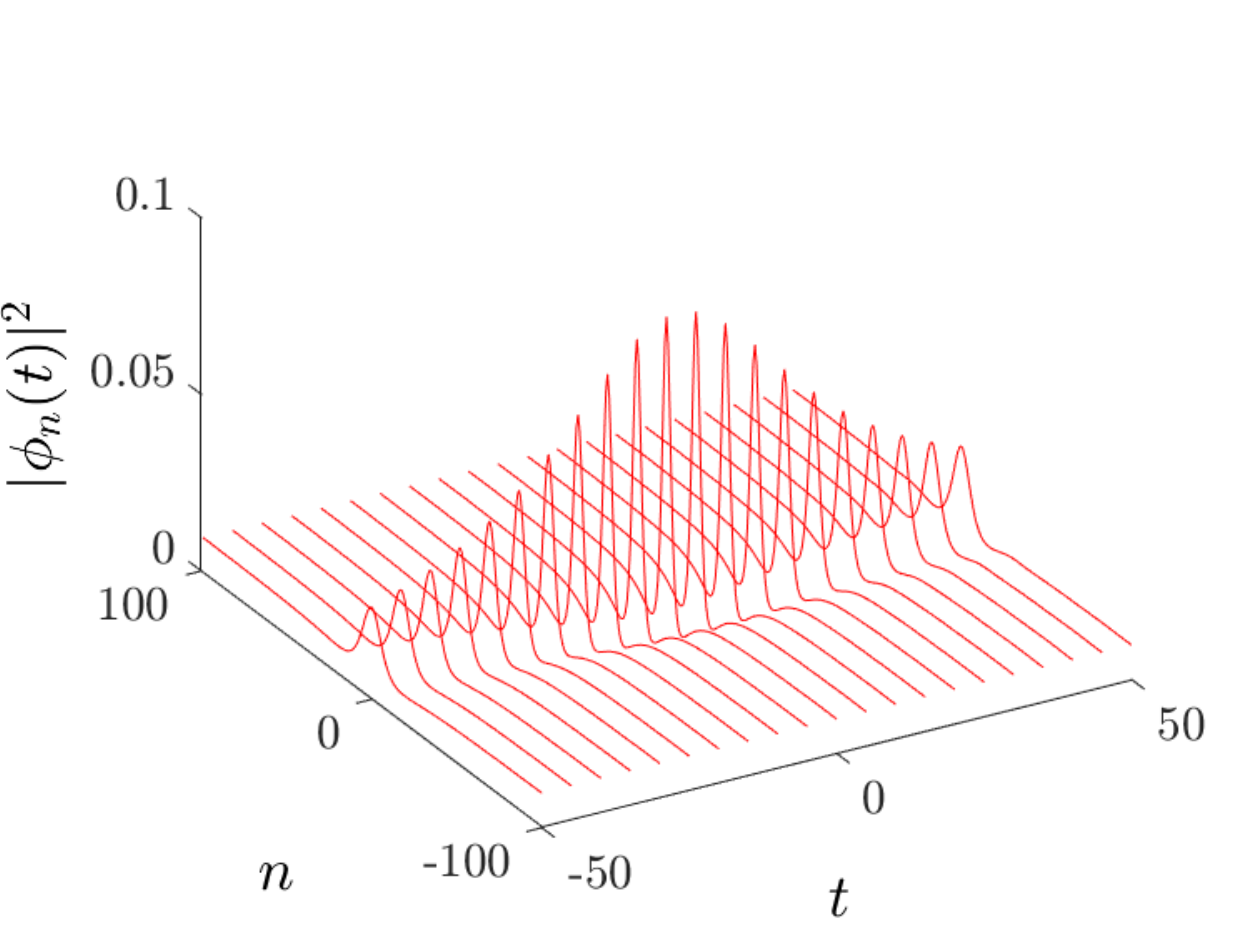}	
		\end{tabular}
	\end{center}
	\caption{Left panel: Evolution of the rational Peregrine initial condition \eqref{per1} with $q=0.1$ in the quintic G-AL lattice \eqref{eq:AL0s} $\sigma=2$, with $\delta=1$, for $K=200$ units and $t\in [-50,50]$. Right panel: The same evolution in the cubic DNLS \eqref{eq:DNLS0s} $\sigma=1$, when $\gamma=1$. Rest of parameters as in the left panel.  Bottom panel: The same evolution in the DNLS with the saturable nonlinearity, when $\gamma=1$. Rest of parameters as in the left panel. }
	\label{fig6}
\end{figure}
\begin{figure}[tbp!]
	\begin{center}
		\begin{tabular}{cc}
			\includegraphics[width=.49\textwidth]{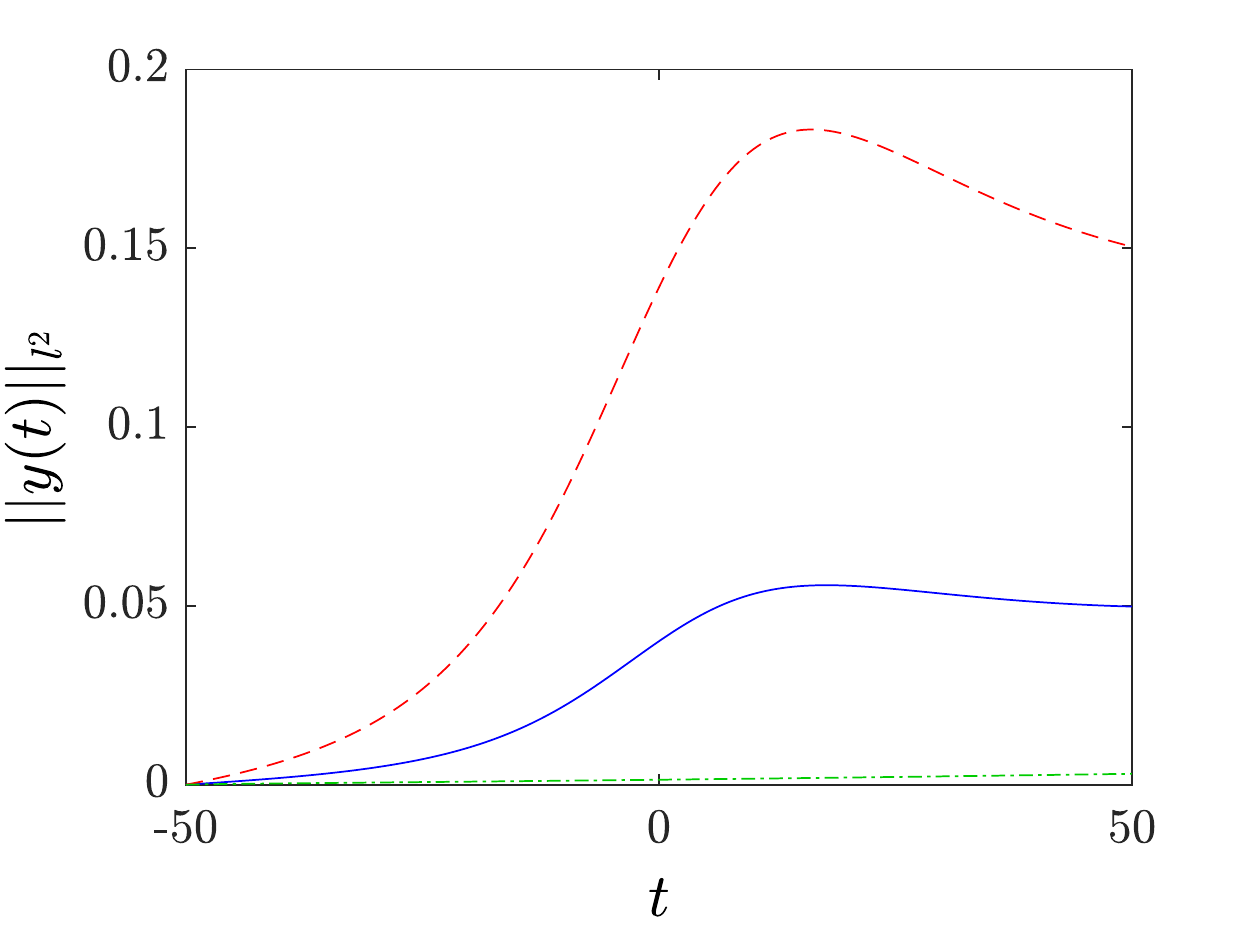}		
			\includegraphics[width=.49\textwidth]{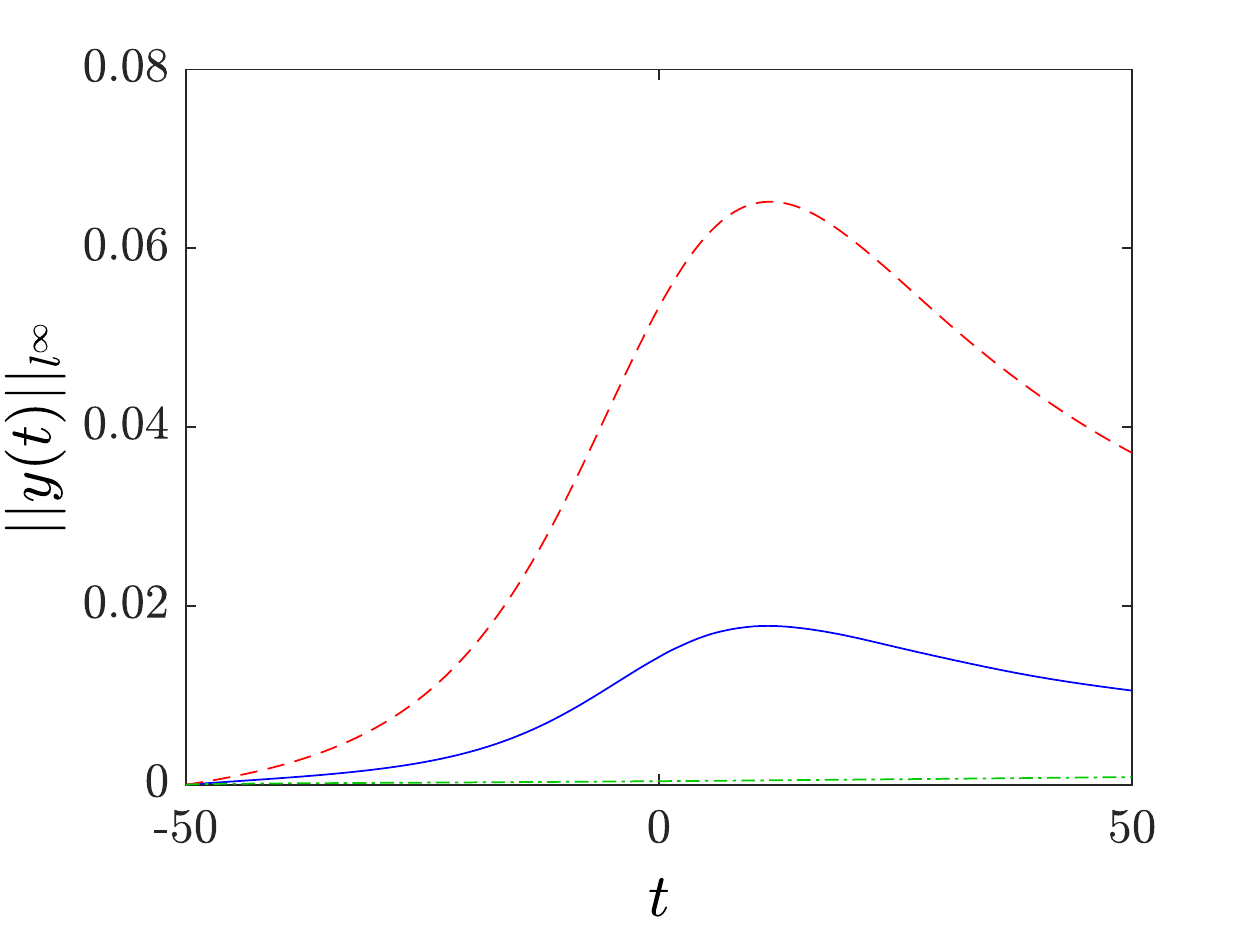}\\
			
		\end{tabular}
	\end{center}
	\caption{Time evolution of $||y(t)||_{l^2}$ (left panel) and $||y(t)||_{l^{\infty}}$ (right panel),  corresponding to the Peregrine soliton dynamics shown in Figure \ref{fig6}, for the saturable DNLS (dashed red curve), the quintic G-AL \eqref{eq:AL0s} (dashed-dotted green curve) and the cubic DNLS (continuous blue curve). }
	\label{fig7}
\end{figure}
%
%
\subsection{Persistence of the AL soliton in the G-AL lattice}
\label{GALsol}
In this section we present a study  of the generalised G-AL lattice \eqref{eq:AL0s} in the cubic case $\sigma=2$ similar to the one presented in \cite{DNJ2021} for the  dynamics of the AL-soliton.
As initial conditions we use the one-soliton solution of the AL \eqref{eq:one-soliton}:
\begin{eqnarray}
&&U_n(0)=\psi^s_n(0)\nonumber\\
\label{sin1}
&=&\frac{\sinh\beta}{\sqrt{\mu}} \mathrm{sech}(\beta n)\exp(i\alpha n),\,\,\,n\in {\mathbb{Z}},\\
&&|| \psi^s(0)||_{l^2}=|| U(0)||_{l^2}=\varepsilon,
\label{sin2}
\end{eqnarray}
where $\alpha \in [-\pi,\pi]$ and $\beta \in [0,\infty)$.  In order to comply with the smallness condition \eqref{sin2} we choose the parameter values accordingly, so that persistence of the corresponding  AL soliton in the DNLS can be expected.

Figure \ref{fig1} depicts the spatio-temporal evolution of the density $|U_n(t)|^2$ of the soliton initial condition when $\mu=1$, $\alpha=\pi/10$ and $\beta=\arcsinh(0.02)$, for the  G-AL equation \eqref{eq:AL0s} with $\delta=1$. The evolution is shown for the time span $t\in [0,2500]$ ($T_{\small{f}}=2500$) and for a chain of $K=2400$ units with periodic boundary conditions.  The evolution of the soliton in the G-AL depicted in the top-left panel of Figure \ref{fig1}  confirms the persistence of the AL soliton with amplitude of order $\mathcal{O}(\varepsilon)$.  Notably, persistence lasts  for a significant large time interval, particularly when one has in mind that our analysis is a ``continuous dependence on the initial data result'' {\em where generally, the time interval of such a dependence on the initial data for a given equation might be short}. Note that for this example of initial condition the value of its $\l^2$-norm  \eqref{sin2} is $\varepsilon=0.2$.

Our numerical results confirm convincingly the analytical predictions presented by the Theorem \ref{Theorem:closeness}, concerning the distance $$y(t)=U(t)-\psi(t),$$
between the solutions of the AL and the G-AL.
The top-right panel of Figure \ref{fig1} depicts the time evolution of $||y(t)||_{l^2}$  and the bottom left panel the evolution of $||y(t)||_{l^{\infty}}$, corresponding to the dynamics of the quintic G-AL shown in the top-left panel of Figure \ref{fig1}. The time evolution of the norm of the distance function  is plotted with a solid line  (green) curve.  Note that since $U(0)=\psi(0)$, the constant $C$ on the estimates \eqref{eq:boundy} and \eqref{eq:boundyNN} reads as 
$$C=2\left(\mu C_{0,\mu}^3+\hat{C}_{0,\sigma,\delta}\right))T_{\tiny{f}}.$$
Since for the considered example of the initial condition $\epsilon\sim\mathcal{O}(10^{-1})$ and the time of integration is of order $T_{\small{f}}\sim \mathcal{O}(10^3)$, the constant $C$ should be of the order $C\sim\mathcal{O}(1)$. Then, according to the analytical estimates, one has for this large $T_{\small{f}}$
\begin{equation*}
	||y(t)||_{l^{\infty}}\leq ||y(t)||_{l^{2}}\leq C\varepsilon^3\sim \mathcal{O}(1),\;\;t\in [0,T_{\small{f}}],
\end{equation*}
The corresponding panels of  Figure \ref{fig1} reveal that the numerical order is significantly lower: $||y(t)||_{l^2}\sim O(10^{-1})$, for $t\in [0,T_{\small{f}}]$. The variation of $||y(t)||_{l^{\infty}}$ is even of smaller order as shown in the bottom panel, showing that $||y(t)||_{l^{\infty}}\sim\mathcal{O}(10^{-2})$.
%

For the numerical time growth-rate of $||y(t)||_{l^2}$ holdsS
\begin{equation*}
	\frac{d}{dt}||y(t)||_{l^2}\sim \mathcal{O}({10^{-4}}),\;\;t\in [0,T_{\small{f}}],
\end{equation*}
which is yet of lower order than the corresponding analytical prediction $\mathcal{O}(10^{-3})$.

The bottom-right panel of Figure \ref{fig1}, depicts the preservation of the soliton's velocity in the G-AL lattice as the corresponding curves  for the G-AL evolution (green curve) and the analytical soliton (black curve), are indistinguishable. For the considered value of $\beta$ we have $\sinh(\beta)\approx\beta$, so that the soliton's speed is $c=2\sin(\alpha)=0.6181$.
%

\subsection{The study of the evolution of the AL-soliton in the DNLS}
\label{qDNLS}
\paragraph{The quintic DNLS.}
The same excellent agreement with the analytical predictions for the persistence of the AL soliton is observed in the case of the generalised DNLS  with the power nonlinearity \ref{pN},
\begin{equation}
i\dot{\phi}_n+{\kappa\nu}\left(\phi_{n+1}-2\phi_n+\phi_{n-1}\right)+\gamma
|\phi_n|^{2\sigma}\phi_n=0,\,\,\,n\in {\mathbb{Z}},\;\;\gamma\in\mathbb{R},\;\;\sigma>0.\label{eq:DNLS0s}
\end{equation}	
The analogous result to Theorem \ref{Theorem:closeness} was proved in \cite{DNJ2021} even for higher dimensional DNLS lattices, which is recalled herein for the sake of completeness.
\begin{theorem}
	\label{ND}
	A. Assume that $\sigma>0$ and that for every $\epsilon>0$,
	the initial conditions of the AL lattice \eqref{eq:AL0} and DNLS equation \eqref{eq:DNLS0s} satisfy
	\begin{eqnarray}
	\label{eq:distance0N}
	|| \phi(0)-\psi(0)||_{l^2}&\le& C_0\, \varepsilon^3,\\
	\label{eq:distance01N}
	|| \phi(0)||_{l^2}&\le& C_{\gamma,0}\,\varepsilon,\\
	\label{eq:Pmu0N}
	P_{\mu}(0)&=&\sum_{n\in\mathbb{Z}^N}\ln(1+\mu|\psi_n(0)|^2)\le \ln(1+(C_{\mu,0}\,\varepsilon)^2),
	\end{eqnarray}
	for some constants $C_0, C_{\gamma,0},\,C_{\mu,0}>0$.	
	Then, for arbitrary $T_{\small{f}}>0$, there exists a constant $C=C(\gamma,\mu,C_{\mu,0},C_{\gamma,0},T_{\small{f}})$, such that the corresponding solutions
	for every $t\in [0,T_{\small{f}}]$, satisfy the estimate 	
	\begin{equation}
	||y(t)||_{l^2}=|| \phi(t)-\psi(t)||_{l^2}\le C\, \varepsilon^3.\label{eq:boundyN}
	\end{equation}
\end{theorem}
B. When $\phi(0)=\psi(0)$, under the assumptions \eqref{eq:distance0N}-\eqref{eq:Pmu0N}, for every $\varepsilon>0$ and $t\in [0,T_{\small{f}}]$, the maximal distance $|| y(t)||_{l^\infty}=\sup_{n\in {\mathbb{Z}^N}}|y_n(t)|=\sup_{n\in {\mathbb{Z}^N}}|\psi_n(t)-\phi_n(t)|$ satisfies the estimate
\begin{equation}
||y(t)||_{l^\infty}\le C \varepsilon^3.\label{eq:boundyhN}
\end{equation}
In Figure \ref{fig2} we present the numerical results for the evolution of the AL-soliton initial condition \eqref{sin1}-\eqref{sin2}, in the quintic DNLS \eqref{eq:DNLS0s} with $\sigma=2$ and $\gamma=1$. We observe that the dynamics are almost identical to the evolution of the generalised G-AL lattice. Still this excellent agreement with the analytical predictions of the associated Theorem \ref{ND} is explained by the very similar functional expression for the the constant $C$ in the estimates \eqref{eq:boundyN} and \eqref{eq:boundyhN} as in Theorem \ref{Theorem:closeness} for the G-AL lattice. This coincidence of the constants explains the same rates of growth in the evolution of the norms of the distance function $y(t)$.

\paragraph{The cubic DNLS: evolution of higher amplitude AL-soliton.} To test the limits of our analytical predictions on the persistence of small amplitude AL solitons we present in this paragraph the results of a numerical study concerning the evolution of the AL-initial condition for a higher initial amplitude corresponding to the case $\beta=\arcsinh(0.1)$   For this choice  of initial condition the value of the $\l^2$-norm  \eqref{sin2} of the initial condition is $\varepsilon=0.4$.

The top panels of Figure \ref{fig3} show the spatiotemporal evolution of the AL-soliton initial condition described above in the cubic DNLS lattice \eqref{eq:DNLS0s} $\sigma=1$ with $\gamma=1$. In the top-left panel the evolution is shown for $t\in [0,2500]$, i.e., $T_f=2500$ and for  a lattice of  $K=2400$ units. The top-right panel shows the evolution for  $t\in [0,1000]$, i.e., $T_f=10000$, for a lattice of $K=7200$ units. We still observe the evolution of a solitonic structure which however, exhibits different dynamical features as the results of the top-right panel reveal: the AL-initial condition evolves as large period-spatially localised breather.

The bottom panels of Figure \ref{fig3} show the evolution of the same AL-soliton initial condition for the quintic DNLS \eqref{eq:DNLS0s} $\sigma=2$ and the remaining parameters are fixed as in the top-panels. In this case the evolution of AL-initial condition exhibits a self-similar decay of its initial amplitude accompanied by an increase of its localisation width. The dynamics tend to achieve a small amplitude solitonic asymptotic state of large width; the solution cannot converge to zero as the system is conservative preserving the initial norm $||\phi(0)||_{l^2}$.

We remark that the observation of the above dynamical features is in agreement with the character of the analytical results based on the continuous dependence of solutions on the initial data: the observed features in Figure \ref{fig3} occur after the finite time interval were the corresponding solutions remain close in profile and amplitude. As it is displayed in Figure \ref{fig3b}, for $t\in [0,50]$ in the cubic case and for $t\in [0,20]$ in the quintic case, the deformations emerging from the initial AL-one soliton in amplitude and profiles are incremental.
  
The above effects can also be found in the time-evolution of the norms of the corresponding distance functions $y(t)$ shown in Figure \ref{fig4}. Note the  time oscillations of the relevant metrics in the case of the cubic DNLS $\sigma=1$ (blue solid line curves) and a decaying oscillatory behavior in the case of the quintic DNLS $\sigma=2$ (dashed red curves).
Remarkably, the analytical estimates are still fulfilled and the illustrated evolution does not contradict the analytical predictions even for large time intervals.  Studying the case $T_f=10000$, which is of order $\mathcal{O}(10^4)$ and for the  initial norm $||\phi(0)||_{l^2}=\epsilon\sim\mathcal{O}(10^{-1})$, the analytical prediction for the upper-bound of the metrics is again satisfied:
In particular, the results depicted in Figure \ref{fig4} support the fact  that the metric  $||y(t)||_{l^2}$ gets altered from order $\mathcal{O}(10^{-1})$ to $\mathcal{O}(1)$ and is of order $\mathcal{O}(10^{-1})$ in the case of  $||y(t)||_{l^{\infty}}$.

As show in Figure \ref{fig5} depicting the evolution of the center of the solitons, the velocity of the AL-soliton is again preserved in large time intervals for the case of $\beta=\arcsinh(0.1)$. For example, in the quintic DNLS of $K=7200$ units the divergence starts at $t\sim 8000$; the persistence of the trace justifies that in the case of the quintic DNLS, the asymptotic state resulting from the initial condition should be  a small amplitude, self-similarly deformed soliton, as explained above.
\subsection{AL-Peregrine soliton in DNLS and generalised AL-systems}
\label{PerDyn}
We conclude the presentation of the numerical results, with what we think is one of the most striking applications of our approach: the persistence of small amplitude rational solutions possessing the functional form of the discrete Peregrine soliton \eqref{prw_exact} of the integrable AL lattice in the non-integrable G-AL and DNLS systems. We present the results for the quintic G-AL \eqref{eq:AL0s} with $\sigma=2$, the cubic DNLS \eqref{eq:DNLS0s} with $\sigma=1$, and the saturable DNLS with the nonlinearity \eqref{satn}; for the saturable DNLS the proof of the variant of Theorem \ref{ND} is also given in \cite[Theorem V.2, pg. 12]{DNJ2021}. We remark that the analytical results on the closeness of solutions can be straightforwardly adapted to cover  the case of the periodic boundary conditions. The latter are suitable to simulate the behavior of solutions on  top of a finite background such as in the case  the AL-PS \eqref{prw_exact}.

For the numerical simulations, since the systems are autonomous and are invariant under the transformation $t\rightarrow -t$, we use as initial condition the AL-analytical Peregrine soliton \eqref{prw_exact}, at some $t=t_0<0$:

\begin{eqnarray}
&&U_n(t_0)=\phi_n(t_0)=\psi^r_n(t_0)\nonumber\\
\label{per1}
&=&q\left[1-\frac{4(1+q^2)(1+4iq^2t_0)}{1+4n^2q^2+16q^4t_0^2(1+q^2)}\right]e^{2iqt_0}.\,\,\,n\in {\mathbb{Z}},\;\;t_0<0.
\end{eqnarray}
We recall that the analytical Peregrine \eqref{prw_exact}  achieves its maximum amplitude at $t=0$. To be in conformity with a smallness condition for the initial amplitude we select $q=0.1$ and set the initial condition at $t_0=-50$.

Figure \ref{fig6} shows the spatiotemporal evolution of the density $|U_n(t)|^2$ for the initial condition \eqref{per1} in the quintic G-AL (left panel) and the cubic DNLS (right panel) while the case of  the saturable DNLS is depicted in the bottom panel. The results demonstrate the persistence of the Peregrine waveforms of small amplitude in all of the non-integrable lattices. In particular, we note that the evolving waveforms  maintain the characteristic spatio-temporal localisation of the discrete Peregrine soliton.

Figure \ref{fig7} portrays the evolution of the norms of the relevant distance functions $y(t)$ confirming once more the closeness analytical results. In the case of the quintic G-AL, the evolution of the norms is plotted with the  dashed-dotted (green) curve, in the case of the cubic DNLS it s plotted with the solid line (blue) curve and in the case of the saturable DNLS with the dashed (red) curve. We observe that in the case of the quintic G-AL, the norms increase incrementally. This should be an effect due to  that both systems have a non-local nonlinearity leading to  that  the solutions remain  closer in comparison with the cases of local nonlinearities. In the case of the cubic DNLS the maximum for both is of order $\mathcal{O}(10^{-2})$. In the case of the saturable DNLS, the maximum of the $l^2$- norm is of order $\mathcal{O}(10^{-1})$,  and of order  $\mathcal{O}(10^{-2})$ for the $l^{\infty}$-norm, fulfilling in an excellent manner the analytical predictions. The relatively larger magnitude variations of the metrics observed for the saturable DNLS can be explained by the specific functional form of the nonlinearity which saturates higher amplitudes. Nevertheless, while  at first-glance these features could be considered  as counter-intuitive for this type of nonlinearity, the analytical results, further corroborated by the numerical studies, confirm that Peregrine type waveforms close to the analytical ones of the integrable AL persist even in the saturable model.

\section{Conclusions}
In this work we have continued our analytical and numerical investigations of the closeness of solutions between the integrable Ablowitz-Ladik lattice and non-integrable lattices for which the Discrete Nonlinear  Schr\"{o}dinger equation serves as the basic underlying model. With our  analytical studies we have established the closeness of solutions, in the sense of a continuous dependence on their initial data, between the  Ablowitz-Ladik system and an important generalisation with an extended power-law nonlinearity. The numerical investigations have explored further the issue of closeness of solutions for the generalised Ablowitz-Ladik system, the Discrete Nonlinear  Schr\"{o}dinger equation with a power-law nonlinearity and with  saturable nonlinearity. For all these examples of  non-integrable systems the numerical studies justified, in full accordance with the analytical predictions, that small amplitude solitonic structures possessing the form of the analytical solutions of the Ablowitz-Ladik system can be excited in nonintegrable lattices and survive for significantly long time intervals. One of the most striking examples is the persistence of spatiotemporal waveforms close to the analytical discrete Peregrine soliton in all of the above mentioned non-integrable lattices.

Future considerations may consider other DNLS systems important for physical applications (such as those relevant for the dynamics of granular crystals \cite{GJ1}), lattices incorporating gain and loss effects \cite{BorisRev}, and crucially,  extensions of our closeness approach to second-order in time lattice dynamical systems (as an example we mention the closeness  between the integrable Toda lattice \cite{Toda1} and Fermi-Pasta-Ulam-Tsingou lattices \cite{FPUT2},\cite{FPUT1}, or Discrete Klein-Gordon systems \cite{DBF}). Such investigations are in progress and results will be reported elsewhere.
\label{secSum}

\end{document}